\documentclass[prl,amsmath,amssymb, notitlepage, twocolumn,
nofootinbib,superscriptaddress,longbibliography]{revtex4-1}
\usepackage{braket}
\usepackage{amsmath,amsfonts, amssymb, amsthm, dsfont}
\usepackage{yfonts}
\usepackage{bm}
\usepackage{mathrsfs}
\usepackage{graphicx}
\usepackage{verbatim}
\usepackage{bbm}
\usepackage{float}

\usepackage{tikz}
\usetikzlibrary{calc}
\usepackage{multirow}

\renewcommand{\vec}[1]{{\mathbf #1}}

\newcommand{\comments}[1]{}

\newcommand{\beq}{\begin{eqnarray}}
\newcommand{\eeq}{\end{eqnarray}}

\newcommand{\Tr}{{\rm Tr}}

\theoremstyle{definition}

\usepackage{color}
\definecolor{darkblue}{rgb}{0.,0.,0.4}
\definecolor{darkred}{rgb}{0.5,0.,0.}
\usepackage[pdftex,colorlinks=true,linkcolor=darkblue,citecolor=blue,urlcolor=darkred]{hyperref}

\newcommand{\beqa}{\begin{equation}\begin{aligned}}
\newcommand{\eeqa}{\end{aligned}\end{equation}}

\renewcommand{\vec}[1]{\bm{#1}}

\graphicspath{{pics/}{}}

\usetikzlibrary{decorations.pathreplacing,decorations.markings}
\tikzset{middlearrow/.style={
		decoration={markings,
			mark= at position 0.55 with {\arrow{#1}} ,
		},
		postaction={decorate}
	}
}

\begin{document}
	
\title{Measurement Protected Quantum Phases}
\author{Shengqi Sang}\affiliation{\PI}
\author{Timothy H. Hsieh}\affiliation{\PI}

\newcommand*{\PI}{Perimeter Institute for Theoretical Physics, Waterloo, Ontario N2L 2Y5, Canada}

	
\begin{abstract}

We introduce a class of hybrid quantum circuits, with random unitaries and projective measurements, which host long-range order in the area law entanglement phase of the steady state.  Our primary example is circuits with unitaries respecting a global Ising symmetry and two competing types of measurements.  The phase diagram has an area law phase with spin glass order, which undergoes a direct transition to a paramagnetic phase with volume law entanglement, as well as a critical regime.  Using mutual information diagnostics, we find that such entanglement transitions preserving a global symmetry are in new universality classes.  We analyze generalizations of such hybrid circuits to higher dimensions, which allow for coexistence of order and volume law entanglement, as well as topological order without any symmetry restrictions.

\end{abstract}
	
\maketitle


A major frontier of quantum many-body physics is understanding what types of order can be stabilized in non-equilibrium settings.
Much progress has stemmed from many-body localization \cite{doi:10.1146/annurev-conmatphys-031214-014726, RevModPhys.91.021001}, which is characterized by the breakdown of thermalization and the restricted growth of entanglement.  Such non-equilibrium features have usually been achieved by considering models with strong disorder, enabling the existence of ``localization protected quantum order'' in highly excited states of Hamiltonians \cite{PhysRevB.88.014206, PhysRevLett.113.107204, bahri}.

Recently, an alternative approach for restricting entanglement growth has been proposed in hybrid quantum circuits involving both unitary evolution and measurements \cite{PhysRevB.98.205136,PhysRevB.100.134306,PhysRevX.9.031009,10.21468/SciPostPhys.7.2.024,PhysRevB.99.224307,PhysRevA.62.062311}.  While generic unitary evolution leads to entanglement growth and a steady state with volume law entanglement, measurements generally disentangle degrees of freedom and lead to area law entangled steady states.  The competition between the two leads to a fascinating phase transition between area and volume law regimes that has been vigorously explored, stimulating alternative perspectives of the entanglement dynamics \cite{choi2019quantum,PhysRevB.100.064204, gullans2019dynamical, PhysRevB.101.104301,PhysRevB.101.104302,PhysRevResearch.2.013022,gullans2019scalable,PhysRevB.101.060301,zhang2020nonuniversal,li2020conformal,fan2020selforganized,shtanko2020classical}.  

A natural question is whether any non-trivial long-range order can be stabilized in the area law regime of such hybrid circuits, and if so, how to understand phase transitions from the ordered phase?  This is motivated by not only the search for new orders and universality classes but also the intriguing possibility of quantum effects being relevant in the brain \cite{FISHER2015593,YUNGERHALPERN201992}.  It has been proposed that the binding of particular molecules realizes projective measurements \cite{FisherE4551}, and thus the possibility of a stable many-body quantum order may be applicable in quantum cognition proposals.     

\begin{figure}
    \includegraphics[width=\linewidth]{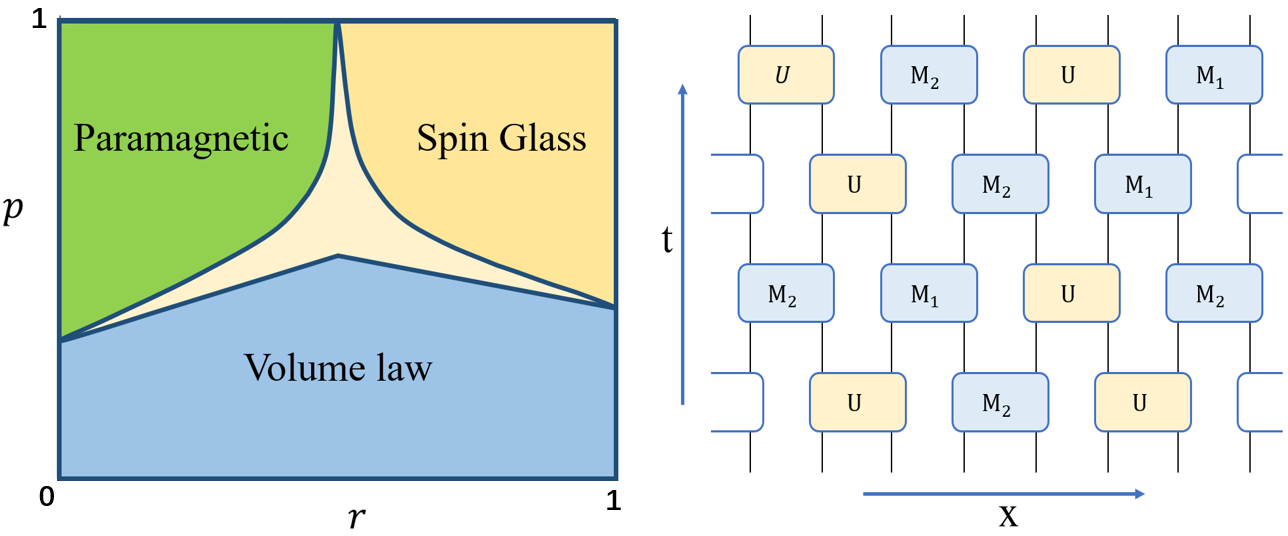}
    \caption{Phase diagram of steady state of hybrid circuit, which consists of brickwall $2$-qubit operations: measurement or $Z_2$-symmetric Clifford unitary with probability $p$ or $1-p$, respectively. Given measurement, it is either $M_1=ZZ$ or $M_2=XI$ measurement with probability $r$ or $1-r$.  The central portion is a critical regime, based on our (finite size $L=768$) numerics, discussed in main text.}
    \label{arch1}
\end{figure}

In this work, we present a class of hybrid circuits which hosts long-range quantum order within the area-law phase.  The basic intuition is that long-range order cannot be connected to a trivial product state by a finite depth circuit \cite{PhysRevLett.97.050401}, and thus the order can survive up to a threshold ratio of unitaries to measurements.   A caveat is that in the circuit architectures presented in \cite{PhysRevB.98.205136,PhysRevB.100.134306,PhysRevX.9.031009}, the basis of any measurement is immediately randomized by a unitary, and we will need to modify the architecture appropriately.  As proof of concept, we demonstrate the existence of long-range spin glass order in the area law phase of a class of hybrid circuits with unitaries respecting a global $Z_2$ symmetry and two competing types of measurements.  We find a phase diagram which contains both a direct transition between spin-glass area law phase and paramagnetic volume law phase as well as a critical regime.  The mutual information at these entanglement transitions exhibits distinct power law scaling, indicating new universality classes due to the global symmetry.  We also analyze a two-dimensional version of the hybrid circuit which enables coexistence of spin-glass order and volume law entanglement. We conclude by mentioning generalizations of our construction to other architectures and stabilizing topological order, which would not require a global symmetry.   

{\bf Setup:} We primarily focus on an ensemble of circuits $C$ acting on a one-dimensional chain of qubits of length $L$ with periodic boundary conditions.  The circuit architecture consists of a brick-wall pattern of two-qubit operations (Fig. \ref{arch1}).  Each operation is either a measurement (M), with probability $p$, or a random unitary (U), with probability $1-p$.  Given that an operation is a measurement, there are two types of measurements $M_1$ and $M_2$, with probability $r$ and  $1-r$.  For two neighboring qubits $i,i+1$, we define $M_1$ to be the projective measurement of $Z_i Z_{i+1}$ and $M_2$ to be the projective measurement of $X_i$.    

As is the case in previous works, it is convenient for scalable simulation to choose the random unitary from an ensemble of Clifford gates, which have the property of mapping a string of Pauli operators to another Pauli string (in the Heisenberg picture).  This enables via the Gottesman-Knill theorem the efficient simulation of the circuit dynamics \cite{PhysRevA.54.1862,gottesman1998heisenberg,PhysRevA.70.052328}, as one need only track the evolution of polynomially many Pauli strings as opposed to an exponentially large wavefunction.  It is important that we add an additional symmetry criteria to this ensemble: each Clifford gate $U$ should map $X_i X_{i+1}$ to itself; this is sufficient and necessary for preserving a global Ising symmetry given by $\prod_{i=1}^L X_i$.  Thus, both unitaries and measurements in the circuit commute with the Ising symmetry, which is clearly essential for defining any symmetry-breaking order.  Details of this ensemble and Clifford/stabilizer technology can be found in the appendix A.  

The initial state is the product state $|\psi_0\rangle = \otimes |+\rangle$, where $X|+\rangle=|+\rangle$.  We are interested in the long time steady state properties after the initial state has been evolved with a deep random circuit $|\psi \rangle  = C|\psi_0\rangle $.  In our simulations, we average target quantities over both different realizations of the circuits and different time slices of a given realization at long time; we hereafter refer to this as ``averaging over the circuit ensemble''.  In particular, to distinguish area and volume law scaling of entanglement, we will compute the Renyi entanglement entropy of $\psi$ after a bipartition into $A$ and $\bar{A}$:
\beqa
S_A = -\log \Tr(\rho_A^2),
\eeqa
averaged over the circuit ensemble.  Here $\rho_A = \Tr_{\bar{A}} |\psi\rangle\langle \psi|$ and as different Renyi entropies are identical for stabilizer states, we have specified without loss of generality the second Renyi. 

We will also compute the spin glass order parameter 
\beqa
O = \frac{1}{L}\sum_{i,j=1}^{L} \langle \psi|Z_i Z_j|\psi \rangle ^2 - \langle\psi|Z_i|\psi \rangle^2 \langle\psi|Z_j|\psi \rangle^2 \label{order}
\eeqa
again averaged over the circuit ensemble.  Given the Ising symmetry $\prod X$, the subtracted piece is always zero.  This order parameter probes long-range entanglement in the following sense.   For a product state, it is manifestly constant (unity), and the application of finite depth circuits can only lead to exponentially decaying correlators $\langle Z_i Z_j\rangle \approx e^{-|i-j|/\xi}$ from Lieb-Robinson bounds \cite{liebrobinson, PhysRevLett.97.050401}.  Hence, in the trivial phase of product-like states, this order parameter is constant (independent of system size). On the other hand, consider an ideal spin glass state: a random cat or Greene-Horne-Zeilinger (GHZ) type state $|s\rangle + (\prod X) |s\rangle$, where $s$ is a random spin configuration in the z-basis; for this state the order parameter grows linearly with $L$ because $\langle Z_i Z_j\rangle^2=1$ for every $i,j$.  Thus, the scaling of this order parameter with system size (constant versus linear) can be used to identify the spin glass phase.  

After averaging, these quantities $S_A, O$ depend only on the parameters of the circuit ensemble $p,r$.  

\begin{figure}
    \includegraphics[width=\linewidth]{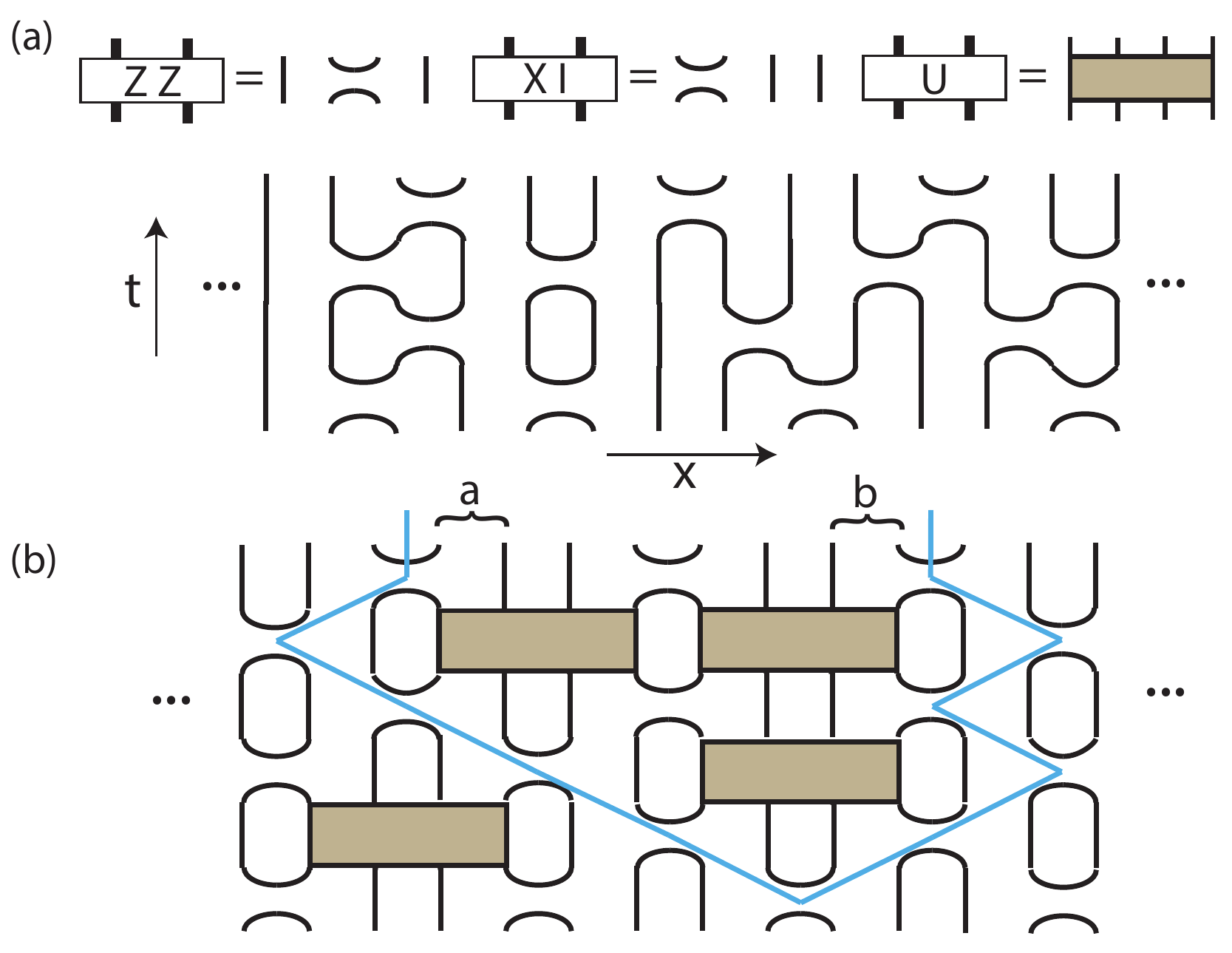}
    \caption{\small (a) The 2-qubit measurements are fermion parity measurements (denoted by the pair of arcs) on the 4 corresponding Majorana modes.  The 2-qubit unitary acts locally on the 4 modes because it preserves fermion parity. A circuit with only measurements maps to loops of Majorana worldlines, as noted in \cite{nahum2019entanglement}. (b) A circuit with only $ZZ$ measurements and unitaries, and the minimal cut (blue line) for an interval with endpoint qubits $(a,b)$.  A minimal cut in the area law phase also mediates spin glass correlation between $a,b$ via the $ZZ$ measurements traversed.}
    \label{majorana}
\end{figure}

{\bf Phase Diagram:} We begin by analyzing several cross-sections of the phase diagram.  

First consider the $p=1$ cross section (circuits with measurement only).  For $r=1$ ($ZZ$ measurement only), the final state has random $Z_i Z_{i+1} = \pm 1$ for each pair of qubits and is thus a random cat state described above (due to the Ising symmetry).  The other extreme $r=0$ yields random paramagnetic product states.  Both spin glass and trivial phases are perturbatively stable to competing measurements respecting the Ising symmetry.  For example, an $X$ measurement on a single qubit $j$ of a cat state will disentangle the qubit and leave the remaining system in a cat state; this is because the stabilizers $Z_{j-1} Z_j, Z_j Z_{j+1}$ become $Z_{j-1} Z_{j+1}, X_j$ after the measurement and the cat heals across $j$. Note that the Ising symmetry is essential here; a $Z$ measurement on a single qubit of a cat state would collapse it into a product state.  

This ensemble of measurement only circuits has a duality between $ZZ, X$ measurements that is manifest after performing a Jordan-Wigner mapping from spins to Majorana modes.  In the latter representation, each spin corresponds to two Majorana modes, and in the resulting Majorana chain, the two types of measurement correspond to fermion parity measurements between pairs of Majoranas on even and odd bonds (Fig. \ref{majorana}a).  The duality fixes a phase transition between spin glass and paramagnetic phases at $r=0.5$, and this critical point in the Majorana representation is explicitly described by a 2d classical loop model at its corresponding critical point; this mapping was detailed in \cite{nahum2019entanglement}.  See Fig. \ref{majorana}a for an example of loops arising from Majorana worldlines. 



\begin{figure}
    \includegraphics[width=\linewidth]{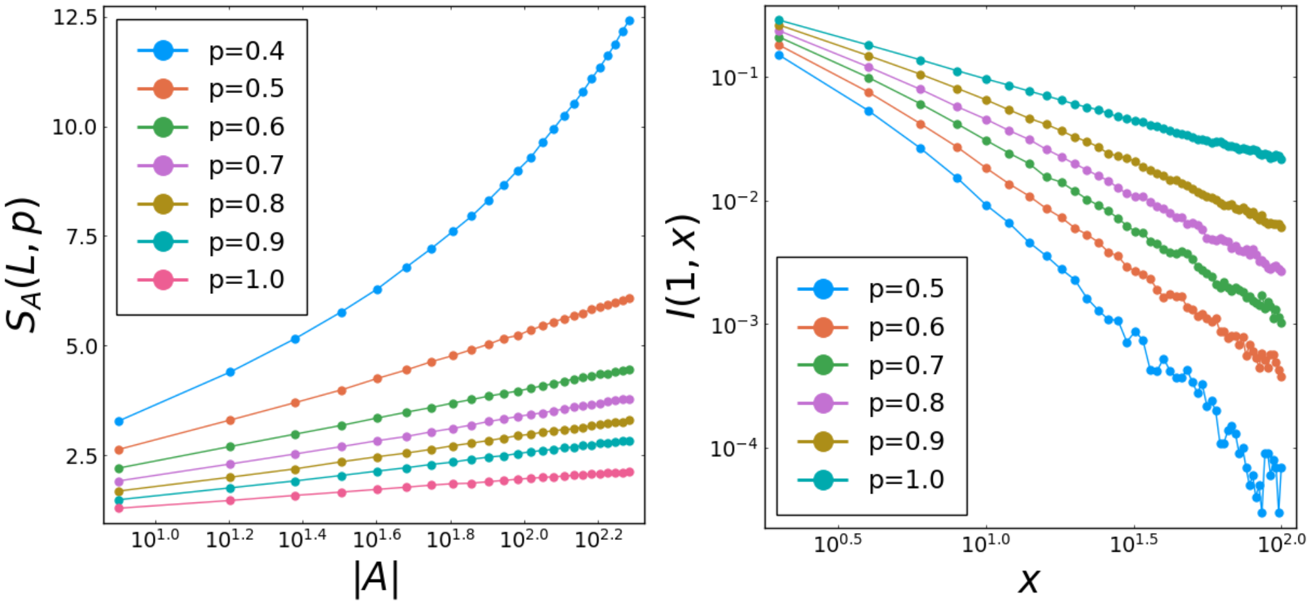}
    \caption{$r=0.5$ cross section (equal probability for $ZZ$ and $X$ measurements). (left) Entanglement entropy versus log of partition size, for various $p$. (right) Mutual information decay in log-log plot. Total system size is $L=768$.}
    \label{50}
\end{figure}

Next, we consider the cross section with fixed $r=0.5$ and variable $p$.  Remarkably, we find (Fig. \ref{50}) that in the range $p\in [0.5,1]$, the entanglement scales with subsystem size as $S_A = c(p) \log |A|$, with coefficient increasing continuously from $c(p=1)\approx 0.27$ (consistent with the loop model prediction $\sqrt{3}/2\pi$ \cite{PhysRevLett.84.3507}).  For $p<0.5$, the entanglement exhibits volume law scaling.  We also compute the mutual information $I(a,b)=S_a+S_b-S_{a\cup b}$ between two qubits $a,b$ and find that in the critical regime, $I$ decays as a power law with $|b-a|$; the power also varies continuously with $p$. 

\begin{figure}
    \includegraphics[width=\linewidth]{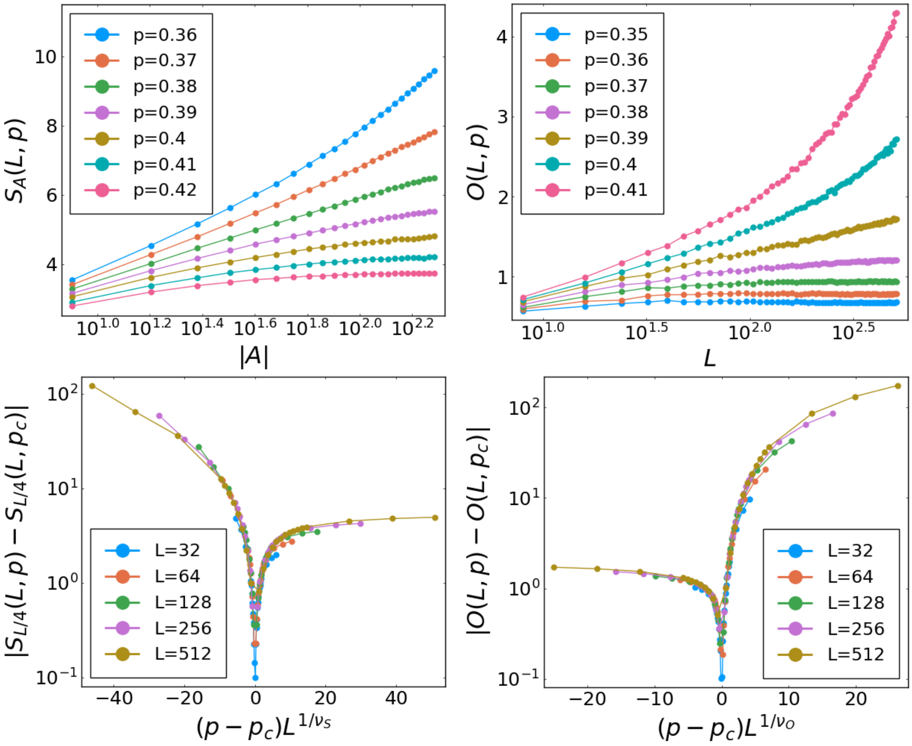}
    \caption{$r=1$ cross section (unitaries and only $ZZ$ measurements with probability $p$). (top) Entanglement entropy versus log of partition size for total system size $L=768$ and spin glass order parameter versus log of system size, for various $p$.  At $p_c\approx0.38,0.39$, the two exhibit log scaling. (bottom) Scaling collapse of both quantities with $\nu_S\approx 1.3, \nu_O \approx 1.5$.}
    \label{100}
\end{figure}

Another important cross section is $r=1.0$ and variable $p$, in which unitaries compete with exclusively $ZZ$ measurements.  We find evidence of a critical point at $p_c \approx 0.38$, in which there is a simultaneous transition from a spin glass area law phase above $p_c$ to a paramagnetic volume law phase below $p_c$.  This is supported by Fig. \ref{100}, which depicts a transition of entanglement scaling from area to volume law at $p_c \approx 0.38$ and a transition of spin glass order parameter from linear scaling with $L$ to constant scaling at $p_c \approx 0.39$ (the two points are within numerical error).  It is evident from the figures that both entanglement and order parameter exhibit log scaling at the critical point and scaling collapse near the critical point: 
\beqa
S_{L/4} (L,p)-S_{L/4} (L,p_c) &=& F\left((p-p_c)L^{1/\nu_S}\right) \\
O (L,p)-O (L,p_c) &=& G\left((p-p_c)L^{1/\nu_O}\right)
\eeqa
with $\nu_S \approx 1.3, \nu_O \approx 1.5$. The critical exponent $\nu_S \approx 1.3$ is comparable to the value ($4/3$) expected for the percolation transition in two dimensions.


Some intuition for this phase transition can be obtained from the Majorana representation, in which a two-qubit unitary acts locally on the four corresponding Majoranas because the unitary respects the Ising symmetry (fermion parity).  In particular, the symmetric 2-qubit Clifford gate is generated by (non-interacting) Majorana swap operations and (interacting) multiplication by the local fermion parity (see Appendix A).  As suggested in \cite{PhysRevX.9.031009}, it is helpful to consider a minimal cut picture which yields the final state's zeroth Renyi entanglement entropy for an interval with endpoints at qubits $a,b$.  We expect the latter to be proportional to the minimum number of unitaries cut by a curve through the circuit with endpoints fixed to be $a,b$ at the final time slice (see Fig. \ref{majorana}b).  Within this picture, the area-law $S_0$ phase corresponds to minimal cuts which pass through a constant number of unitaries as $|b-a| \rightarrow \infty$. Though the minimal cut picture is only strictly valid for the zeroth Renyi entropy of a circuit with Haar random unitaries, we use it as a heuristic for understanding the transition in the Clifford circuit.    


Such a minimal cut in the area law phase also implies that the spin glass correlation $\langle Z_a Z_b \rangle^2$ is constant as $|b-a| \rightarrow \infty$, yielding a long-range spin glass.  This arises from the product of $ZZ$s from the measurements along the minimal cut, which is attenuated by only a constant number of unitaries traversed by the cut.  The $ZZ$ correlation begins with the bottom two qubits of the minimal cut, and as the next measurements along the minimal cut are performed, the pair of qubits which are correlated propagates outward in both directions until it reaches $a,b$. 


Hence, the minimal cut links the area law phase and spin glass order, at least for $r=1$. A minimal cut through only a constant number of unitaries is no longer possible when the unitary cluster percolates.  Hence, we expect that the area law spin glass is destroyed near the site percolation threshold of the square lattice $0.59$.
Our numerical results  indicates $1-p_c \approx 0.62$, which is close to the value $0.59$ given by the minimum cut picture.


\begin{figure}
    \includegraphics[width=\linewidth]{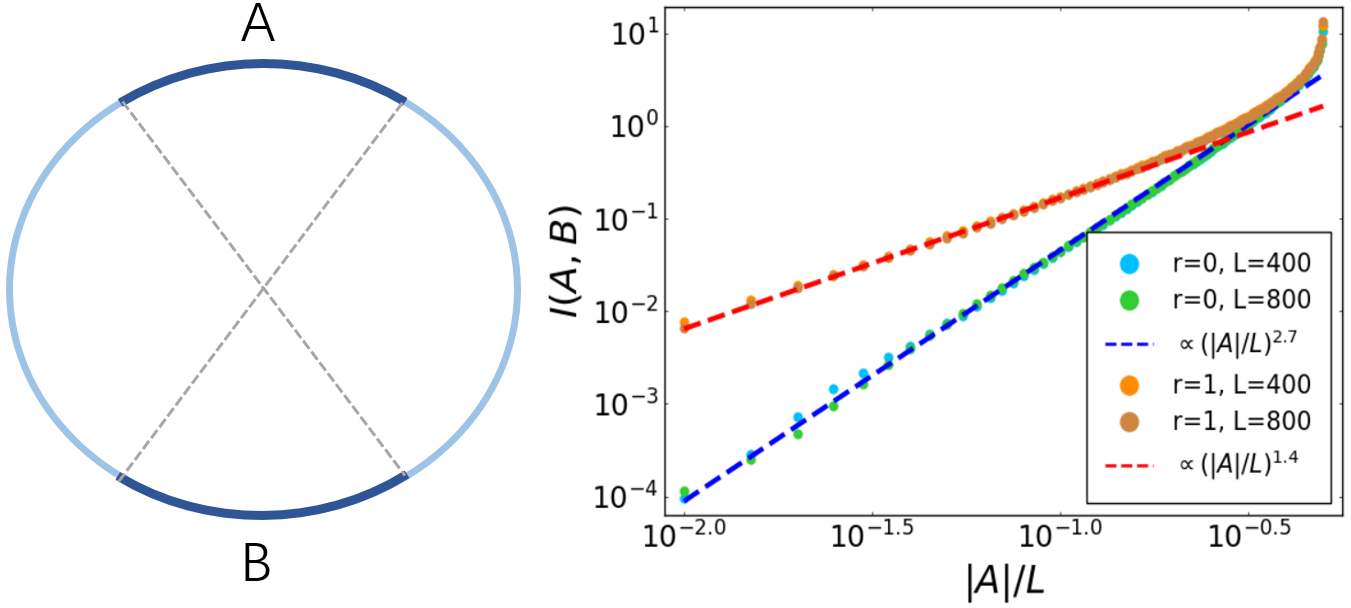}
    \caption{(left) Antipodal geometry. Intervals $A$ and $B$ are of the same size and centered on two antipodal points of the periodic qubit chain of length $L$. (right) Simulated mutual information $I(A,B)$ as a function of the ratio $|A|/L$ at critical points corresponding to $r=0$ ($Z_2$-Clifford unitaries + $X$ measurement) and $r=1$ ($Z_2$-Clifford unitaries + $ZZ$ measurement).}
    \label{conformal}
\end{figure}

A useful probe of the critical point is the mutual information between two antipodal intervals $A,B$ of equal size $|A|$ (see Fig. \ref{conformal}).  In previous studies without a global symmetry, including both Haar and Clifford random circuits, the mutual information scales as $I_{A,B} \propto (|A|/L)^4$ \cite{PhysRevB.100.134306,PhysRevX.9.031009} in the regime of $(|A|/L) \ll 1$.  In contrast, in our symmetric hybrid circuit, at the spin glass area-law to paramagnetic volume-law transition described above, we find $I_{A,B} \propto (|A|/L)^{1.4}$.  Moreover, in the $r=0$ cross section (which involves $X$ measurement only), there is a direct transition between paramagnetic area-law and paramagnetic volume-law, at which we find $I_{A,B} \propto (|A|/L)^{2.7}$ (Fig. \ref{conformal}).  These indicate that the entanglement transitions in the presence of a global symmetry are in distinct universality classes than those without symmetry.

The full phase diagram is presented in Fig. 1 and obtained from both cross sections presented above and additional ones in the Appendix C ($r=0.25,0.75$ and $p=0.75$).  The shaded central portion is a critical regime including the segment of the $r=0.5$ cross section discussed earlier, with logarithmic entanglement scaling in our current system sizes.  One possibility is that the segment at $r=0.5$ closely borders two phase boundaries and thus appears critical in finite systems.  However, both the large range of the log scaling observed ($p\in[0.5,1]$) as well as the sharp transition from log to volume law scaling (Fig. \ref{50}) (as opposed to a smooth crossover) are surprising.  In short, the critical phase in the thermodynamic limit either (1) persists as a critical phase (2) disappears, resulting in two critical lines separating volume law from paramagnetic and spin glass or (3) disappears, resulting in three critical lines meeting at a point.     

For understanding the critical regime, one may consider loop models with crossings \cite{PhysRevB.87.184204,PhysRevLett.90.090601,READ2001409,crossingloop,PhysRevA.44.2410,PhysRevLett.81.504,Kager_2006,Ikhlef_2007} as toy models for our hybrid circuit.  As mentioned, the $r=0.5,p=1$ critical point is described by non-intersecting loops, and loop crossings represent unitaries which swap Majoranas; these serve as bottlenecks in the circuit/loop configuration which lengthen the minimal cut and increase entanglement.  Interestingly, for finite crossing probability, loop models have a critical ``Goldstone phase'' \cite{PhysRevB.87.184204,PhysRevLett.90.090601,READ2001409,crossingloop,PhysRevA.44.2410,PhysRevLett.81.504,Kager_2006,Ikhlef_2007}, referring to a sigma model description in the continuum.  Indeed, the phase diagram in \cite{PhysRevB.87.184204} bears much similarity to ours, and it would be interesting to understand in detail the connection.  This Goldstone phase has also been discussed in the context of entanglement transitions in random tensor networks \cite{vasseur}.    

\begin{figure}
    \includegraphics[width=\linewidth]{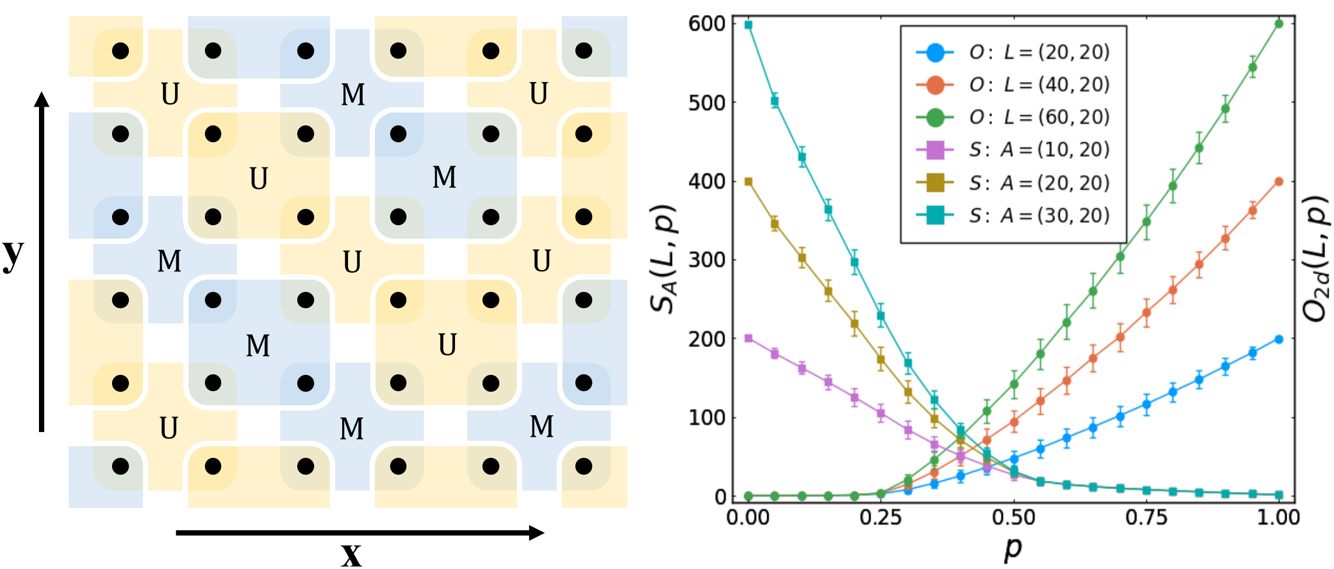}
    \caption{(left) A cross section of the $(2+1)$d circuit architecture. Black dots are qubits and each colored square denotes an operation acting on qubits within the square. An operation is either a random Ising-symmetric 4-qubit Clifford gate (yellow) or a series of 3 two-qubit measurements $Z_1 Z_2, Z_2 Z_3, Z_3 Z_4$ (blue), with probability $1-p$ or $p$. The circuit alternates between the background and foreground, which consist of two distinct partitions into sets of four. (right) The two order parameters $O(L, p)$ and $S_A(L, p)$ as functions of $p$. For the simulation of $S_A (L,p)$ the total system size is fixed to be $L = (L_x, L_y) = (60, 20)$.}
    \label{2d}
\end{figure}

{\bf Higher dimensions:} In contrast to one dimension, higher-dimensional circuit architectures allow for the possibility that both measurement and unitary clusters can percolate in a parameter range.  In such a range, the final state consists of an extensive subset of spins connected in the past by a measurement cluster, enabling spin-glass order.  On the other hand, the percolating unitary cluster may intersect a minimal surface an extensive number of times, leading to volume law entanglement scaling. 

This can be verified by simulating a generalization of our architecture to a two-dimensional system using the generalized order parameter
\beqa
O_{2d}(L, p) = \frac{1}{L_x L_y}\sum_{\vec{i},\vec{j}} \langle \psi|Z_{\vec{i}} Z_{\vec{j}}|\psi \rangle ^2 - \langle\psi|Z_{\vec{i}}|\psi \rangle^2 \langle\psi|Z_{\vec{j}}|\psi \rangle^2
\eeqa
The left panel of Fig.\ref{2d} shows a temporal cross-section of our $(2+1)d$ circuit; the circuit alternates between the background and foreground, which involve two distinct partitions into sets of 4-qubit operations. Each operation is, with probability $1-p$ or $p$, either a random 4-qubit Clifford gate commuting with $X_1 X_2 X_3 X_4$, or 3 consecutive measurements in bases $Z_1 Z_2, Z_2 Z_3$ and $Z_3 Z_4$. Our result (right panel of Fig. \ref{2d}) shows that the entanglement transition and the spin glass transition respectively happen at $p_{c, S} \approx 0.5$ and $p_{c,O} \approx 0.25$. The parameter range between them corresponds to a volume-law, spin-glass-ordered phase.  See Appendix E for more details.

\begin{figure}
    \includegraphics[width=\linewidth]{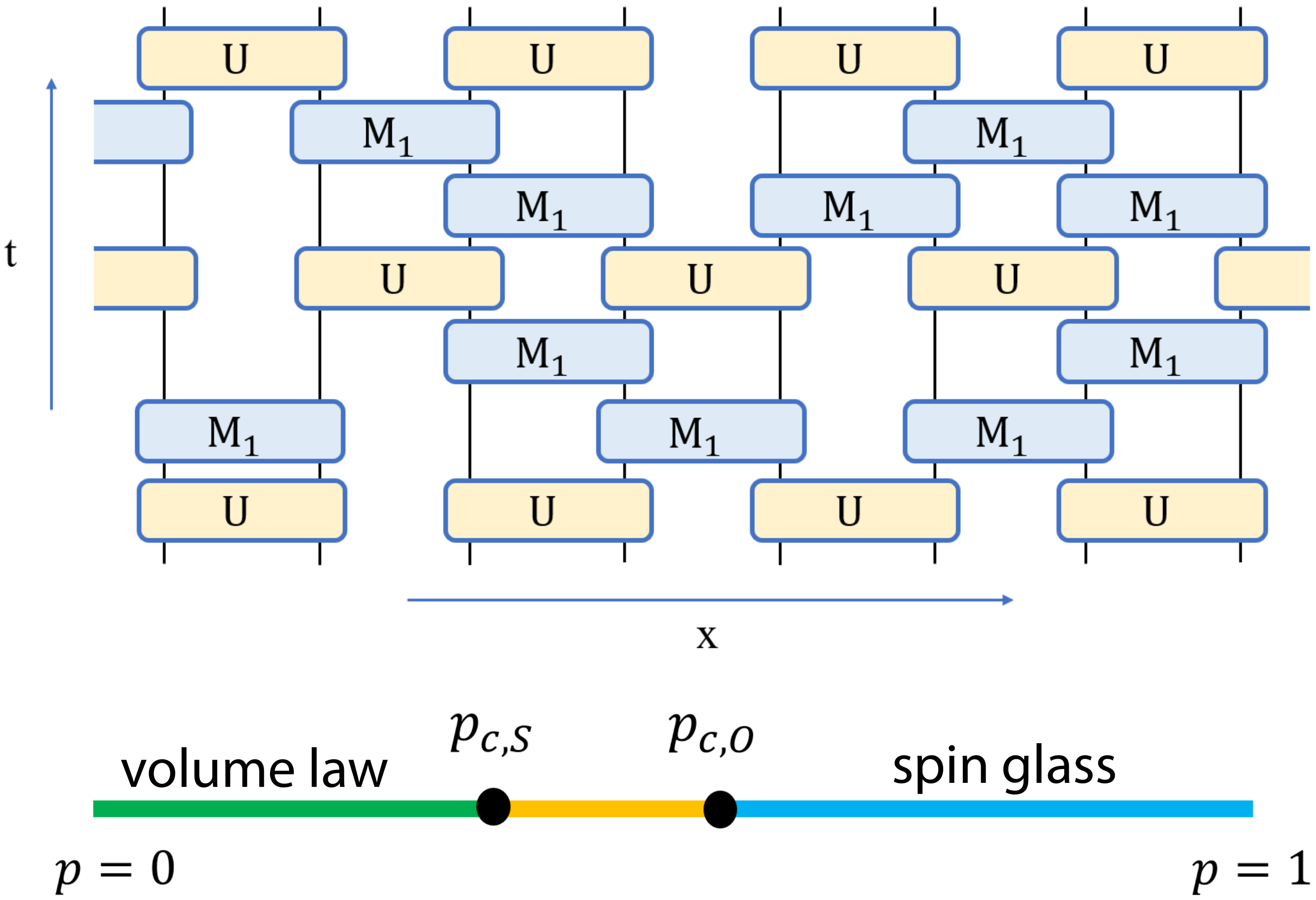}
    \caption{Alternative class of circuits with $Z_2$-symmetric random unitaries always applied in brick-wall fashion, with in-between layers of $M_1=Z_i Z_{i+1}$ measurements for every $i$, each applied with probability $p$.  Phase diagram has a paramagnetic volume law phase, a critical regime, and spin glass area law phase.  $p_{c,S}\approx 0.52,p_{c,O}\approx 0.58, $ (see Appendix D for details).}
    \label{arch2}
\end{figure}


{\bf Other architectures:} We expect that replacing symmetric Clifford with symmetric Haar random unitaries will not change the qualitative aspects of the phase diagrams.  The stability of the ordered phase derives from the inability of finite depth circuits to destroy the order, and this holds for any symmetric unitary circuit.  

Furthermore, the hybrid circuits considered in the literature can also support spin glass order, after a small but important modification.  Such circuits are brick walls of operations that have probability $1-p$ of being a random unitary and probability $p$ of being a projective measurement followed by a random unitary, and previous work has considered this setup with $ZZ$ measurements as the projective measurement \cite{PhysRevB.98.205136}.  However, both the fact that a unitary is {\it always} applied, even after a measurement is made, and the fact that each measurement layer was restricted to either even bonds or odd bonds of qubits \cite{PhysRevB.98.205136}, implies that the measurement basis is irrelevant and no spin-glass order can exist.  

Consider a very similar setup in which random unitaries are always applied in a brick wall pattern, but between each layer of unitaries, $ZZ$ measurements on {\it any} neighboring qubits are performed with probability $p$ (see Fig. \ref{arch2}).  In this case, for large $p$, connected clusters of $ZZ$ measurements are performed, and the subsequent (single) layer of unitaries cannot destroy the spin-glass order as long as the Lieb-Robinson length is shorter than the typical measurement cluster size.  As before, it is essential that each random unitary respect the global Ising symmetry.   

We find that the volume law phase persists up to $p \approx  0.52$ and the spin glass order begins at $p \approx  0.58$ (see Appendix D for data).  Based on our numerics, the interval between these points appears to be critical (entanglement scaling is not strictly area or volume law), but we cannot rule out finite size effects masking a direct transition between the two phases.  Nonetheless, the existence of the spin glass phase is unambiguous.  
  
{\bf Topological order and beyond:}  The measurements in the hybrid circuit can be generalized to stabilize other types of quantum order.  For example, measurements of the (commuting Pauli) operators in the toric code Hamiltonian \cite{KITAEV20062} would stabilize a random topologically ordered state.  One could use the same type of order parameter as Eq. (\ref{order}), with the $Z$ operator replaced by a string operator (one of the Wilson lines).  In contrast to the spin glass order, the unitaries in this hybrid circuit need not respect any global symmetry for the stability of topological order.

The use of measurements to protect against random operations also forms the basis of active quantum error correction.  An important difference is that active quantum error correction seeks to reverse errors by applying operations depending on the error syndromes are obtained.  In contrast, in our setup, while the measurement operations are essential, their outcomes are not important (our scheme has no feedback).  This is because the protocol does not preserve a particular quantum state but instead a particular long-range entanglement structure.  Hence the name measurement protected quantum {\it phases}.

Nonetheless, it would be interesting to incorporate conditioning on measurements.  More generally, can the hybrid unitary and measurement circuits lead to new quantum orders beyond the critical points?  And how can these new universality classes with symmetry be understood?  We leave these explorations for the future.    

{\it Note added}: While completing this work, we noticed a related work \cite{lavasani2020measurementinduced} posted recently.

{\it Acknowledgements.}
We thank X. Chen, M.P.A. Fisher, Y. Li, and A. Nahum for useful discussions, as well as R. Melko for computational resources. The authors acknowledge support from the Natural Sciences and Engineering Research Council of Canada (NSERC) through a
Discovery Grant. Research at Perimeter Institute is supported in part by the Government of Canada through the Department of Innovation, Science and Economic Development Canada and by the Province of Ontario through the Ministry of Colleges and Universities.  This work
was made possible by the facilities of the Shared Hierarchical Academic Research Computing Network (SHARCNET) and Compute/Calcul Canada.

\bibliography{main.bib}

\begin{thebibliography}{44}%
\makeatletter
\providecommand \@ifxundefined [1]{%
 \@ifx{#1\undefined}
}%
\providecommand \@ifnum [1]{%
 \ifnum #1\expandafter \@firstoftwo
 \else \expandafter \@secondoftwo
 \fi
}%
\providecommand \@ifx [1]{%
 \ifx #1\expandafter \@firstoftwo
 \else \expandafter \@secondoftwo
 \fi
}%
\providecommand \natexlab [1]{#1}%
\providecommand \enquote  [1]{``#1''}%
\providecommand \bibnamefont  [1]{#1}%
\providecommand \bibfnamefont [1]{#1}%
\providecommand \citenamefont [1]{#1}%
\providecommand \href@noop [0]{\@secondoftwo}%
\providecommand \href [0]{\begingroup \@sanitize@url \@href}%
\providecommand \@href[1]{\@@startlink{#1}\@@href}%
\providecommand \@@href[1]{\endgroup#1\@@endlink}%
\providecommand \@sanitize@url [0]{\catcode `\\12\catcode `\$12\catcode
  `\&12\catcode `\#12\catcode `\^12\catcode `\_12\catcode `\%12\relax}%
\providecommand \@@startlink[1]{}%
\providecommand \@@endlink[0]{}%
\providecommand \url  [0]{\begingroup\@sanitize@url \@url }%
\providecommand \@url [1]{\endgroup\@href {#1}{\urlprefix }}%
\providecommand \urlprefix  [0]{URL }%
\providecommand \Eprint [0]{\href }%
\providecommand \doibase [0]{http://dx.doi.org/}%
\providecommand \selectlanguage [0]{\@gobble}%
\providecommand \bibinfo  [0]{\@secondoftwo}%
\providecommand \bibfield  [0]{\@secondoftwo}%
\providecommand \translation [1]{[#1]}%
\providecommand \BibitemOpen [0]{}%
\providecommand \bibitemStop [0]{}%
\providecommand \bibitemNoStop [0]{.\EOS\space}%
\providecommand \EOS [0]{\spacefactor3000\relax}%
\providecommand \BibitemShut  [1]{\csname bibitem#1\endcsname}%
\let\auto@bib@innerbib\@empty
\bibitem [{\citenamefont {Nandkishore}\ and\ \citenamefont
  {Huse}(2015)}]{doi:10.1146/annurev-conmatphys-031214-014726}%
  \BibitemOpen
  \bibfield  {author} {\bibinfo {author} {\bibfnamefont {Rahul}\ \bibnamefont
  {Nandkishore}}\ and\ \bibinfo {author} {\bibfnamefont {David~A.}\
  \bibnamefont {Huse}},\ }\bibfield  {title} {\enquote {\bibinfo {title}
  {Many-body localization and thermalization in quantum statistical
  mechanics},}\ }\href {\doibase 10.1146/annurev-conmatphys-031214-014726}
  {\bibfield  {journal} {\bibinfo  {journal} {Annual Review of Condensed Matter
  Physics}\ }\textbf {\bibinfo {volume} {6}},\ \bibinfo {pages} {15--38}
  (\bibinfo {year} {2015})},\ \Eprint
  {http://arxiv.org/abs/https://doi.org/10.1146/annurev-conmatphys-031214-014726}
  {https://doi.org/10.1146/annurev-conmatphys-031214-014726} \BibitemShut
  {NoStop}%
\bibitem [{\citenamefont {Abanin}\ \emph {et~al.}(2019)\citenamefont {Abanin},
  \citenamefont {Altman}, \citenamefont {Bloch},\ and\ \citenamefont
  {Serbyn}}]{RevModPhys.91.021001}%
  \BibitemOpen
  \bibfield  {author} {\bibinfo {author} {\bibfnamefont {Dmitry~A.}\
  \bibnamefont {Abanin}}, \bibinfo {author} {\bibfnamefont {Ehud}\ \bibnamefont
  {Altman}}, \bibinfo {author} {\bibfnamefont {Immanuel}\ \bibnamefont
  {Bloch}}, \ and\ \bibinfo {author} {\bibfnamefont {Maksym}\ \bibnamefont
  {Serbyn}},\ }\bibfield  {title} {\enquote {\bibinfo {title} {Colloquium:
  Many-body localization, thermalization, and entanglement},}\ }\href {\doibase
  10.1103/RevModPhys.91.021001} {\bibfield  {journal} {\bibinfo  {journal}
  {Rev. Mod. Phys.}\ }\textbf {\bibinfo {volume} {91}},\ \bibinfo {pages}
  {021001} (\bibinfo {year} {2019})}\BibitemShut {NoStop}%
\bibitem [{\citenamefont {Huse}\ \emph {et~al.}(2013)\citenamefont {Huse},
  \citenamefont {Nandkishore}, \citenamefont {Oganesyan}, \citenamefont {Pal},\
  and\ \citenamefont {Sondhi}}]{PhysRevB.88.014206}%
  \BibitemOpen
  \bibfield  {author} {\bibinfo {author} {\bibfnamefont {David~A.}\
  \bibnamefont {Huse}}, \bibinfo {author} {\bibfnamefont {Rahul}\ \bibnamefont
  {Nandkishore}}, \bibinfo {author} {\bibfnamefont {Vadim}\ \bibnamefont
  {Oganesyan}}, \bibinfo {author} {\bibfnamefont {Arijeet}\ \bibnamefont
  {Pal}}, \ and\ \bibinfo {author} {\bibfnamefont {S.~L.}\ \bibnamefont
  {Sondhi}},\ }\bibfield  {title} {\enquote {\bibinfo {title}
  {Localization-protected quantum order},}\ }\href {\doibase
  10.1103/PhysRevB.88.014206} {\bibfield  {journal} {\bibinfo  {journal} {Phys.
  Rev. B}\ }\textbf {\bibinfo {volume} {88}},\ \bibinfo {pages} {014206}
  (\bibinfo {year} {2013})}\BibitemShut {NoStop}%
\bibitem [{\citenamefont {Kj\"all}\ \emph {et~al.}(2014)\citenamefont
  {Kj\"all}, \citenamefont {Bardarson},\ and\ \citenamefont
  {Pollmann}}]{PhysRevLett.113.107204}%
  \BibitemOpen
  \bibfield  {author} {\bibinfo {author} {\bibfnamefont {Jonas~A.}\
  \bibnamefont {Kj\"all}}, \bibinfo {author} {\bibfnamefont {Jens~H.}\
  \bibnamefont {Bardarson}}, \ and\ \bibinfo {author} {\bibfnamefont {Frank}\
  \bibnamefont {Pollmann}},\ }\bibfield  {title} {\enquote {\bibinfo {title}
  {Many-body localization in a disordered quantum ising chain},}\ }\href
  {\doibase 10.1103/PhysRevLett.113.107204} {\bibfield  {journal} {\bibinfo
  {journal} {Phys. Rev. Lett.}\ }\textbf {\bibinfo {volume} {113}},\ \bibinfo
  {pages} {107204} (\bibinfo {year} {2014})}\BibitemShut {NoStop}%
\bibitem [{\citenamefont {Bahri}\ \emph {et~al.}(2015)\citenamefont {Bahri},
  \citenamefont {Vosk}, \citenamefont {Altman},\ and\ \citenamefont
  {Vishwanath}}]{bahri}%
  \BibitemOpen
  \bibfield  {author} {\bibinfo {author} {\bibfnamefont {Yasaman}\ \bibnamefont
  {Bahri}}, \bibinfo {author} {\bibfnamefont {Ronen}\ \bibnamefont {Vosk}},
  \bibinfo {author} {\bibfnamefont {Ehud}\ \bibnamefont {Altman}}, \ and\
  \bibinfo {author} {\bibfnamefont {Ashvin}\ \bibnamefont {Vishwanath}},\
  }\bibfield  {title} {\enquote {\bibinfo {title} {Localization and topology
  protected quantum coherence at the edge of hot matter},}\ }\href {\doibase
  10.1038/ncomms8341} {\bibfield  {journal} {\bibinfo  {journal} {Nature
  Communications}\ }\textbf {\bibinfo {volume} {6}},\ \bibinfo {pages} {7341}
  (\bibinfo {year} {2015})}\BibitemShut {NoStop}%
\bibitem [{\citenamefont {Li}\ \emph {et~al.}(2018)\citenamefont {Li},
  \citenamefont {Chen},\ and\ \citenamefont {Fisher}}]{PhysRevB.98.205136}%
  \BibitemOpen
  \bibfield  {author} {\bibinfo {author} {\bibfnamefont {Yaodong}\ \bibnamefont
  {Li}}, \bibinfo {author} {\bibfnamefont {Xiao}\ \bibnamefont {Chen}}, \ and\
  \bibinfo {author} {\bibfnamefont {Matthew P.~A.}\ \bibnamefont {Fisher}},\
  }\bibfield  {title} {\enquote {\bibinfo {title} {Quantum zeno effect and the
  many-body entanglement transition},}\ }\href {\doibase
  10.1103/PhysRevB.98.205136} {\bibfield  {journal} {\bibinfo  {journal} {Phys.
  Rev. B}\ }\textbf {\bibinfo {volume} {98}},\ \bibinfo {pages} {205136}
  (\bibinfo {year} {2018})}\BibitemShut {NoStop}%
\bibitem [{\citenamefont {Li}\ \emph {et~al.}(2019)\citenamefont {Li},
  \citenamefont {Chen},\ and\ \citenamefont {Fisher}}]{PhysRevB.100.134306}%
  \BibitemOpen
  \bibfield  {author} {\bibinfo {author} {\bibfnamefont {Yaodong}\ \bibnamefont
  {Li}}, \bibinfo {author} {\bibfnamefont {Xiao}\ \bibnamefont {Chen}}, \ and\
  \bibinfo {author} {\bibfnamefont {Matthew P.~A.}\ \bibnamefont {Fisher}},\
  }\bibfield  {title} {\enquote {\bibinfo {title} {Measurement-driven
  entanglement transition in hybrid quantum circuits},}\ }\href {\doibase
  10.1103/PhysRevB.100.134306} {\bibfield  {journal} {\bibinfo  {journal}
  {Phys. Rev. B}\ }\textbf {\bibinfo {volume} {100}},\ \bibinfo {pages}
  {134306} (\bibinfo {year} {2019})}\BibitemShut {NoStop}%
\bibitem [{\citenamefont {Skinner}\ \emph {et~al.}(2019)\citenamefont
  {Skinner}, \citenamefont {Ruhman},\ and\ \citenamefont
  {Nahum}}]{PhysRevX.9.031009}%
  \BibitemOpen
  \bibfield  {author} {\bibinfo {author} {\bibfnamefont {Brian}\ \bibnamefont
  {Skinner}}, \bibinfo {author} {\bibfnamefont {Jonathan}\ \bibnamefont
  {Ruhman}}, \ and\ \bibinfo {author} {\bibfnamefont {Adam}\ \bibnamefont
  {Nahum}},\ }\bibfield  {title} {\enquote {\bibinfo {title}
  {Measurement-induced phase transitions in the dynamics of entanglement},}\
  }\href {\doibase 10.1103/PhysRevX.9.031009} {\bibfield  {journal} {\bibinfo
  {journal} {Phys. Rev. X}\ }\textbf {\bibinfo {volume} {9}},\ \bibinfo {pages}
  {031009} (\bibinfo {year} {2019})}\BibitemShut {NoStop}%
\bibitem [{\citenamefont {Cao}\ \emph {et~al.}(2019)\citenamefont {Cao},
  \citenamefont {Tilloy},\ and\ \citenamefont
  {Luca}}]{10.21468/SciPostPhys.7.2.024}%
  \BibitemOpen
  \bibfield  {author} {\bibinfo {author} {\bibfnamefont {Xiangyu}\ \bibnamefont
  {Cao}}, \bibinfo {author} {\bibfnamefont {Antoine}\ \bibnamefont {Tilloy}}, \
  and\ \bibinfo {author} {\bibfnamefont {Andrea~De}\ \bibnamefont {Luca}},\
  }\bibfield  {title} {\enquote {\bibinfo {title} {{Entanglement in a fermion
  chain under continuous monitoring}},}\ }\href {\doibase
  10.21468/SciPostPhys.7.2.024} {\bibfield  {journal} {\bibinfo  {journal}
  {SciPost Phys.}\ }\textbf {\bibinfo {volume} {7}},\ \bibinfo {pages} {24}
  (\bibinfo {year} {2019})}\BibitemShut {NoStop}%
\bibitem [{\citenamefont {Chan}\ \emph {et~al.}(2019)\citenamefont {Chan},
  \citenamefont {Nandkishore}, \citenamefont {Pretko},\ and\ \citenamefont
  {Smith}}]{PhysRevB.99.224307}%
  \BibitemOpen
  \bibfield  {author} {\bibinfo {author} {\bibfnamefont {Amos}\ \bibnamefont
  {Chan}}, \bibinfo {author} {\bibfnamefont {Rahul~M.}\ \bibnamefont
  {Nandkishore}}, \bibinfo {author} {\bibfnamefont {Michael}\ \bibnamefont
  {Pretko}}, \ and\ \bibinfo {author} {\bibfnamefont {Graeme}\ \bibnamefont
  {Smith}},\ }\bibfield  {title} {\enquote {\bibinfo {title}
  {Unitary-projective entanglement dynamics},}\ }\href {\doibase
  10.1103/PhysRevB.99.224307} {\bibfield  {journal} {\bibinfo  {journal} {Phys.
  Rev. B}\ }\textbf {\bibinfo {volume} {99}},\ \bibinfo {pages} {224307}
  (\bibinfo {year} {2019})}\BibitemShut {NoStop}%
\bibitem [{\citenamefont {Aharonov}(2000)}]{PhysRevA.62.062311}%
  \BibitemOpen
  \bibfield  {author} {\bibinfo {author} {\bibfnamefont {Dorit}\ \bibnamefont
  {Aharonov}},\ }\bibfield  {title} {\enquote {\bibinfo {title} {Quantum to
  classical phase transition in noisy quantum computers},}\ }\href {\doibase
  10.1103/PhysRevA.62.062311} {\bibfield  {journal} {\bibinfo  {journal} {Phys.
  Rev. A}\ }\textbf {\bibinfo {volume} {62}},\ \bibinfo {pages} {062311}
  (\bibinfo {year} {2000})}\BibitemShut {NoStop}%
\bibitem [{\citenamefont {Choi}\ \emph {et~al.}(2019)\citenamefont {Choi},
  \citenamefont {Bao}, \citenamefont {Qi},\ and\ \citenamefont
  {Altman}}]{choi2019quantum}%
  \BibitemOpen
  \bibfield  {author} {\bibinfo {author} {\bibfnamefont {Soonwon}\ \bibnamefont
  {Choi}}, \bibinfo {author} {\bibfnamefont {Yimu}\ \bibnamefont {Bao}},
  \bibinfo {author} {\bibfnamefont {Xiao-Liang}\ \bibnamefont {Qi}}, \ and\
  \bibinfo {author} {\bibfnamefont {Ehud}\ \bibnamefont {Altman}},\ }\href@noop
  {} {\enquote {\bibinfo {title} {Quantum error correction in scrambling
  dynamics and measurement induced phase transition},}\ } (\bibinfo {year}
  {2019}),\ \Eprint {http://arxiv.org/abs/1903.05124} {arXiv:1903.05124
  [quant-ph]} \BibitemShut {NoStop}%
\bibitem [{\citenamefont {Szyniszewski}\ \emph {et~al.}(2019)\citenamefont
  {Szyniszewski}, \citenamefont {Romito},\ and\ \citenamefont
  {Schomerus}}]{PhysRevB.100.064204}%
  \BibitemOpen
  \bibfield  {author} {\bibinfo {author} {\bibfnamefont {M.}~\bibnamefont
  {Szyniszewski}}, \bibinfo {author} {\bibfnamefont {A.}~\bibnamefont
  {Romito}}, \ and\ \bibinfo {author} {\bibfnamefont {H.}~\bibnamefont
  {Schomerus}},\ }\bibfield  {title} {\enquote {\bibinfo {title} {Entanglement
  transition from variable-strength weak measurements},}\ }\href {\doibase
  10.1103/PhysRevB.100.064204} {\bibfield  {journal} {\bibinfo  {journal}
  {Phys. Rev. B}\ }\textbf {\bibinfo {volume} {100}},\ \bibinfo {pages}
  {064204} (\bibinfo {year} {2019})}\BibitemShut {NoStop}%
\bibitem [{\citenamefont {Gullans}\ and\ \citenamefont
  {Huse}(2019{\natexlab{a}})}]{gullans2019dynamical}%
  \BibitemOpen
  \bibfield  {author} {\bibinfo {author} {\bibfnamefont {Michael~J.}\
  \bibnamefont {Gullans}}\ and\ \bibinfo {author} {\bibfnamefont {David~A.}\
  \bibnamefont {Huse}},\ }\href@noop {} {\enquote {\bibinfo {title} {Dynamical
  purification phase transition induced by quantum measurements},}\ } (\bibinfo
  {year} {2019}{\natexlab{a}}),\ \Eprint {http://arxiv.org/abs/1905.05195}
  {arXiv:1905.05195 [quant-ph]} \BibitemShut {NoStop}%
\bibitem [{\citenamefont {Bao}\ \emph {et~al.}(2020)\citenamefont {Bao},
  \citenamefont {Choi},\ and\ \citenamefont {Altman}}]{PhysRevB.101.104301}%
  \BibitemOpen
  \bibfield  {author} {\bibinfo {author} {\bibfnamefont {Yimu}\ \bibnamefont
  {Bao}}, \bibinfo {author} {\bibfnamefont {Soonwon}\ \bibnamefont {Choi}}, \
  and\ \bibinfo {author} {\bibfnamefont {Ehud}\ \bibnamefont {Altman}},\
  }\bibfield  {title} {\enquote {\bibinfo {title} {Theory of the phase
  transition in random unitary circuits with measurements},}\ }\href {\doibase
  10.1103/PhysRevB.101.104301} {\bibfield  {journal} {\bibinfo  {journal}
  {Phys. Rev. B}\ }\textbf {\bibinfo {volume} {101}},\ \bibinfo {pages}
  {104301} (\bibinfo {year} {2020})}\BibitemShut {NoStop}%
\bibitem [{\citenamefont {Jian}\ \emph {et~al.}(2020)\citenamefont {Jian},
  \citenamefont {You}, \citenamefont {Vasseur},\ and\ \citenamefont
  {Ludwig}}]{PhysRevB.101.104302}%
  \BibitemOpen
  \bibfield  {author} {\bibinfo {author} {\bibfnamefont {Chao-Ming}\
  \bibnamefont {Jian}}, \bibinfo {author} {\bibfnamefont {Yi-Zhuang}\
  \bibnamefont {You}}, \bibinfo {author} {\bibfnamefont {Romain}\ \bibnamefont
  {Vasseur}}, \ and\ \bibinfo {author} {\bibfnamefont {Andreas W.~W.}\
  \bibnamefont {Ludwig}},\ }\bibfield  {title} {\enquote {\bibinfo {title}
  {Measurement-induced criticality in random quantum circuits},}\ }\href
  {\doibase 10.1103/PhysRevB.101.104302} {\bibfield  {journal} {\bibinfo
  {journal} {Phys. Rev. B}\ }\textbf {\bibinfo {volume} {101}},\ \bibinfo
  {pages} {104302} (\bibinfo {year} {2020})}\BibitemShut {NoStop}%
\bibitem [{\citenamefont {Tang}\ and\ \citenamefont
  {Zhu}(2020)}]{PhysRevResearch.2.013022}%
  \BibitemOpen
  \bibfield  {author} {\bibinfo {author} {\bibfnamefont {Qicheng}\ \bibnamefont
  {Tang}}\ and\ \bibinfo {author} {\bibfnamefont {W.}~\bibnamefont {Zhu}},\
  }\bibfield  {title} {\enquote {\bibinfo {title} {Measurement-induced phase
  transition: A case study in the nonintegrable model by density-matrix
  renormalization group calculations},}\ }\href {\doibase
  10.1103/PhysRevResearch.2.013022} {\bibfield  {journal} {\bibinfo  {journal}
  {Phys. Rev. Research}\ }\textbf {\bibinfo {volume} {2}},\ \bibinfo {pages}
  {013022} (\bibinfo {year} {2020})}\BibitemShut {NoStop}%
\bibitem [{\citenamefont {Gullans}\ and\ \citenamefont
  {Huse}(2019{\natexlab{b}})}]{gullans2019scalable}%
  \BibitemOpen
  \bibfield  {author} {\bibinfo {author} {\bibfnamefont {Michael~J.}\
  \bibnamefont {Gullans}}\ and\ \bibinfo {author} {\bibfnamefont {David~A.}\
  \bibnamefont {Huse}},\ }\href@noop {} {\enquote {\bibinfo {title} {Scalable
  probes of measurement-induced criticality},}\ } (\bibinfo {year}
  {2019}{\natexlab{b}}),\ \Eprint {http://arxiv.org/abs/1910.00020}
  {arXiv:1910.00020 [cond-mat.stat-mech]} \BibitemShut {NoStop}%
\bibitem [{\citenamefont {Zabalo}\ \emph {et~al.}(2020)\citenamefont {Zabalo},
  \citenamefont {Gullans}, \citenamefont {Wilson}, \citenamefont
  {Gopalakrishnan}, \citenamefont {Huse},\ and\ \citenamefont
  {Pixley}}]{PhysRevB.101.060301}%
  \BibitemOpen
  \bibfield  {author} {\bibinfo {author} {\bibfnamefont {Aidan}\ \bibnamefont
  {Zabalo}}, \bibinfo {author} {\bibfnamefont {Michael~J.}\ \bibnamefont
  {Gullans}}, \bibinfo {author} {\bibfnamefont {Justin~H.}\ \bibnamefont
  {Wilson}}, \bibinfo {author} {\bibfnamefont {Sarang}\ \bibnamefont
  {Gopalakrishnan}}, \bibinfo {author} {\bibfnamefont {David~A.}\ \bibnamefont
  {Huse}}, \ and\ \bibinfo {author} {\bibfnamefont {J.~H.}\ \bibnamefont
  {Pixley}},\ }\bibfield  {title} {\enquote {\bibinfo {title} {Critical
  properties of the measurement-induced transition in random quantum
  circuits},}\ }\href {\doibase 10.1103/PhysRevB.101.060301} {\bibfield
  {journal} {\bibinfo  {journal} {Phys. Rev. B}\ }\textbf {\bibinfo {volume}
  {101}},\ \bibinfo {pages} {060301} (\bibinfo {year} {2020})}\BibitemShut
  {NoStop}%
\bibitem [{\citenamefont {Zhang}\ \emph {et~al.}(2020)\citenamefont {Zhang},
  \citenamefont {Reyes}, \citenamefont {Kourtis}, \citenamefont {Chamon},
  \citenamefont {Mucciolo},\ and\ \citenamefont
  {Ruckenstein}}]{zhang2020nonuniversal}%
  \BibitemOpen
  \bibfield  {author} {\bibinfo {author} {\bibfnamefont {Lei}\ \bibnamefont
  {Zhang}}, \bibinfo {author} {\bibfnamefont {Justin~A.}\ \bibnamefont
  {Reyes}}, \bibinfo {author} {\bibfnamefont {Stefanos}\ \bibnamefont
  {Kourtis}}, \bibinfo {author} {\bibfnamefont {Claudio}\ \bibnamefont
  {Chamon}}, \bibinfo {author} {\bibfnamefont {Eduardo~R.}\ \bibnamefont
  {Mucciolo}}, \ and\ \bibinfo {author} {\bibfnamefont {Andrei~E.}\
  \bibnamefont {Ruckenstein}},\ }\href@noop {} {\enquote {\bibinfo {title}
  {Nonuniversal entanglement level statistics in projection-driven quantum
  circuits},}\ } (\bibinfo {year} {2020}),\ \Eprint
  {http://arxiv.org/abs/2001.11428} {arXiv:2001.11428 [cond-mat.stat-mech]}
  \BibitemShut {NoStop}%
\bibitem [{\citenamefont {Li}\ \emph {et~al.}(2020)\citenamefont {Li},
  \citenamefont {Chen}, \citenamefont {Ludwig},\ and\ \citenamefont
  {Fisher}}]{li2020conformal}%
  \BibitemOpen
  \bibfield  {author} {\bibinfo {author} {\bibfnamefont {Yaodong}\ \bibnamefont
  {Li}}, \bibinfo {author} {\bibfnamefont {Xiao}\ \bibnamefont {Chen}},
  \bibinfo {author} {\bibfnamefont {Andreas W.~W.}\ \bibnamefont {Ludwig}}, \
  and\ \bibinfo {author} {\bibfnamefont {Matthew P.~A.}\ \bibnamefont
  {Fisher}},\ }\href@noop {} {\enquote {\bibinfo {title} {Conformal invariance
  and quantum non-locality in hybrid quantum circuits},}\ } (\bibinfo {year}
  {2020}),\ \Eprint {http://arxiv.org/abs/2003.12721} {arXiv:2003.12721
  [quant-ph]} \BibitemShut {NoStop}%
\bibitem [{\citenamefont {Fan}\ \emph {et~al.}(2020)\citenamefont {Fan},
  \citenamefont {Vijay}, \citenamefont {Vishwanath},\ and\ \citenamefont
  {You}}]{fan2020selforganized}%
  \BibitemOpen
  \bibfield  {author} {\bibinfo {author} {\bibfnamefont {Ruihua}\ \bibnamefont
  {Fan}}, \bibinfo {author} {\bibfnamefont {Sagar}\ \bibnamefont {Vijay}},
  \bibinfo {author} {\bibfnamefont {Ashvin}\ \bibnamefont {Vishwanath}}, \ and\
  \bibinfo {author} {\bibfnamefont {Yi-Zhuang}\ \bibnamefont {You}},\
  }\href@noop {} {\enquote {\bibinfo {title} {Self-organized error correction
  in random unitary circuits with measurement},}\ } (\bibinfo {year} {2020}),\
  \Eprint {http://arxiv.org/abs/2002.12385} {arXiv:2002.12385
  [cond-mat.stat-mech]} \BibitemShut {NoStop}%
\bibitem [{\citenamefont {Shtanko}\ \emph {et~al.}(2020)\citenamefont
  {Shtanko}, \citenamefont {Kharkov}, \citenamefont {García-Pintos},\ and\
  \citenamefont {Gorshkov}}]{shtanko2020classical}%
  \BibitemOpen
  \bibfield  {author} {\bibinfo {author} {\bibfnamefont {Oles}\ \bibnamefont
  {Shtanko}}, \bibinfo {author} {\bibfnamefont {Yaroslav~A.}\ \bibnamefont
  {Kharkov}}, \bibinfo {author} {\bibfnamefont {Luis~Pedro}\ \bibnamefont
  {García-Pintos}}, \ and\ \bibinfo {author} {\bibfnamefont {Alexey~V.}\
  \bibnamefont {Gorshkov}},\ }\href@noop {} {\enquote {\bibinfo {title}
  {Classical models of entanglement in monitored random circuits},}\ }
  (\bibinfo {year} {2020}),\ \Eprint {http://arxiv.org/abs/2004.06736}
  {arXiv:2004.06736 [cond-mat.dis-nn]} \BibitemShut {NoStop}%
\bibitem [{\citenamefont {Fisher}(2015)}]{FISHER2015593}%
  \BibitemOpen
  \bibfield  {author} {\bibinfo {author} {\bibfnamefont {Matthew~P.A.}\
  \bibnamefont {Fisher}},\ }\bibfield  {title} {\enquote {\bibinfo {title}
  {Quantum cognition: The possibility of processing with nuclear spins in the
  brain},}\ }\href {\doibase https://doi.org/10.1016/j.aop.2015.08.020}
  {\bibfield  {journal} {\bibinfo  {journal} {Annals of Physics}\ }\textbf
  {\bibinfo {volume} {362}},\ \bibinfo {pages} {593 -- 602} (\bibinfo {year}
  {2015})}\BibitemShut {NoStop}%
\bibitem [{\citenamefont {Halpern]}\ and\ \citenamefont
  {Crosson}(2019)}]{YUNGERHALPERN201992}%
  \BibitemOpen
  \bibfield  {author} {\bibinfo {author} {\bibfnamefont {Nicole~[Yunger}\
  \bibnamefont {Halpern]}}\ and\ \bibinfo {author} {\bibfnamefont {Elizabeth}\
  \bibnamefont {Crosson}},\ }\bibfield  {title} {\enquote {\bibinfo {title}
  {Quantum information in the posner model of quantum cognition},}\ }\href
  {\doibase https://doi.org/10.1016/j.aop.2018.11.016} {\bibfield  {journal}
  {\bibinfo  {journal} {Annals of Physics}\ }\textbf {\bibinfo {volume}
  {407}},\ \bibinfo {pages} {92 -- 147} (\bibinfo {year} {2019})}\BibitemShut
  {NoStop}%
\bibitem [{\citenamefont {Fisher}\ and\ \citenamefont
  {Radzihovsky}(2018)}]{FisherE4551}%
  \BibitemOpen
  \bibfield  {author} {\bibinfo {author} {\bibfnamefont {Matthew P.~A.}\
  \bibnamefont {Fisher}}\ and\ \bibinfo {author} {\bibfnamefont {Leo}\
  \bibnamefont {Radzihovsky}},\ }\bibfield  {title} {\enquote {\bibinfo {title}
  {Quantum indistinguishability in chemical reactions},}\ }\href {\doibase
  10.1073/pnas.1718402115} {\bibfield  {journal} {\bibinfo  {journal}
  {Proceedings of the National Academy of Sciences}\ }\textbf {\bibinfo
  {volume} {115}},\ \bibinfo {pages} {E4551--E4558} (\bibinfo {year} {2018})},\
  \Eprint
  {http://arxiv.org/abs/https://www.pnas.org/content/115/20/E4551.full.pdf}
  {https://www.pnas.org/content/115/20/E4551.full.pdf} \BibitemShut {NoStop}%
\bibitem [{\citenamefont {Bravyi}\ \emph {et~al.}(2006)\citenamefont {Bravyi},
  \citenamefont {Hastings},\ and\ \citenamefont
  {Verstraete}}]{PhysRevLett.97.050401}%
  \BibitemOpen
  \bibfield  {author} {\bibinfo {author} {\bibfnamefont {S.}~\bibnamefont
  {Bravyi}}, \bibinfo {author} {\bibfnamefont {M.~B.}\ \bibnamefont
  {Hastings}}, \ and\ \bibinfo {author} {\bibfnamefont {F.}~\bibnamefont
  {Verstraete}},\ }\bibfield  {title} {\enquote {\bibinfo {title}
  {Lieb-robinson bounds and the generation of correlations and topological
  quantum order},}\ }\href {\doibase 10.1103/PhysRevLett.97.050401} {\bibfield
  {journal} {\bibinfo  {journal} {Phys. Rev. Lett.}\ }\textbf {\bibinfo
  {volume} {97}},\ \bibinfo {pages} {050401} (\bibinfo {year}
  {2006})}\BibitemShut {NoStop}%
\bibitem [{\citenamefont {Gottesman}(1996)}]{PhysRevA.54.1862}%
  \BibitemOpen
  \bibfield  {author} {\bibinfo {author} {\bibfnamefont {Daniel}\ \bibnamefont
  {Gottesman}},\ }\bibfield  {title} {\enquote {\bibinfo {title} {Class of
  quantum error-correcting codes saturating the quantum hamming bound},}\
  }\href {\doibase 10.1103/PhysRevA.54.1862} {\bibfield  {journal} {\bibinfo
  {journal} {Phys. Rev. A}\ }\textbf {\bibinfo {volume} {54}},\ \bibinfo
  {pages} {1862--1868} (\bibinfo {year} {1996})}\BibitemShut {NoStop}%
\bibitem [{\citenamefont {Gottesman}(1998)}]{gottesman1998heisenberg}%
  \BibitemOpen
  \bibfield  {author} {\bibinfo {author} {\bibfnamefont {Daniel}\ \bibnamefont
  {Gottesman}},\ }\href@noop {} {\enquote {\bibinfo {title} {The heisenberg
  representation of quantum computers},}\ } (\bibinfo {year} {1998}),\ \Eprint
  {http://arxiv.org/abs/quant-ph/9807006} {arXiv:quant-ph/9807006 [quant-ph]}
  \BibitemShut {NoStop}%
\bibitem [{\citenamefont {Aaronson}\ and\ \citenamefont
  {Gottesman}(2004)}]{PhysRevA.70.052328}%
  \BibitemOpen
  \bibfield  {author} {\bibinfo {author} {\bibfnamefont {Scott}\ \bibnamefont
  {Aaronson}}\ and\ \bibinfo {author} {\bibfnamefont {Daniel}\ \bibnamefont
  {Gottesman}},\ }\bibfield  {title} {\enquote {\bibinfo {title} {Improved
  simulation of stabilizer circuits},}\ }\href {\doibase
  10.1103/PhysRevA.70.052328} {\bibfield  {journal} {\bibinfo  {journal} {Phys.
  Rev. A}\ }\textbf {\bibinfo {volume} {70}},\ \bibinfo {pages} {052328}
  (\bibinfo {year} {2004})}\BibitemShut {NoStop}%
\bibitem [{\citenamefont {Lieb}\ and\ \citenamefont
  {Robinson}(1972)}]{liebrobinson}%
  \BibitemOpen
  \bibfield  {author} {\bibinfo {author} {\bibfnamefont {Elliott~H.}\
  \bibnamefont {Lieb}}\ and\ \bibinfo {author} {\bibfnamefont {Derek~W.}\
  \bibnamefont {Robinson}},\ }\bibfield  {title} {\enquote {\bibinfo {title}
  {The finite group velocity of quantum spin systems},}\ }\href {\doibase
  10.1007/BF01645779} {\bibfield  {journal} {\bibinfo  {journal}
  {Communications in Mathematical Physics}\ }\textbf {\bibinfo {volume} {28}},\
  \bibinfo {pages} {251--257} (\bibinfo {year} {1972})}\BibitemShut {NoStop}%
\bibitem [{\citenamefont {Nahum}\ and\ \citenamefont
  {Skinner}(2019)}]{nahum2019entanglement}%
  \BibitemOpen
  \bibfield  {author} {\bibinfo {author} {\bibfnamefont {Adam}\ \bibnamefont
  {Nahum}}\ and\ \bibinfo {author} {\bibfnamefont {Brian}\ \bibnamefont
  {Skinner}},\ }\href@noop {} {\enquote {\bibinfo {title} {Entanglement and
  dynamics of diffusion-annihilation processes with majorana defects},}\ }
  (\bibinfo {year} {2019}),\ \Eprint {http://arxiv.org/abs/1911.11169}
  {arXiv:1911.11169 [cond-mat.stat-mech]} \BibitemShut {NoStop}%
\bibitem [{\citenamefont {Cardy}(2000)}]{PhysRevLett.84.3507}%
  \BibitemOpen
  \bibfield  {author} {\bibinfo {author} {\bibfnamefont {John}\ \bibnamefont
  {Cardy}},\ }\bibfield  {title} {\enquote {\bibinfo {title} {Linking numbers
  for self-avoiding loops and percolation: Application to the spin quantum hall
  transition},}\ }\href {\doibase 10.1103/PhysRevLett.84.3507} {\bibfield
  {journal} {\bibinfo  {journal} {Phys. Rev. Lett.}\ }\textbf {\bibinfo
  {volume} {84}},\ \bibinfo {pages} {3507--3510} (\bibinfo {year}
  {2000})}\BibitemShut {NoStop}%
\bibitem [{\citenamefont {Nahum}\ \emph {et~al.}(2013)\citenamefont {Nahum},
  \citenamefont {Serna}, \citenamefont {Somoza},\ and\ \citenamefont
  {Ortu\~no}}]{PhysRevB.87.184204}%
  \BibitemOpen
  \bibfield  {author} {\bibinfo {author} {\bibfnamefont {Adam}\ \bibnamefont
  {Nahum}}, \bibinfo {author} {\bibfnamefont {P.}~\bibnamefont {Serna}},
  \bibinfo {author} {\bibfnamefont {A.~M.}\ \bibnamefont {Somoza}}, \ and\
  \bibinfo {author} {\bibfnamefont {M.}~\bibnamefont {Ortu\~no}},\ }\bibfield
  {title} {\enquote {\bibinfo {title} {Loop models with crossings},}\ }\href
  {\doibase 10.1103/PhysRevB.87.184204} {\bibfield  {journal} {\bibinfo
  {journal} {Phys. Rev. B}\ }\textbf {\bibinfo {volume} {87}},\ \bibinfo
  {pages} {184204} (\bibinfo {year} {2013})}\BibitemShut {NoStop}%
\bibitem [{\citenamefont {Jacobsen}\ \emph {et~al.}(2003)\citenamefont
  {Jacobsen}, \citenamefont {Read},\ and\ \citenamefont
  {Saleur}}]{PhysRevLett.90.090601}%
  \BibitemOpen
  \bibfield  {author} {\bibinfo {author} {\bibfnamefont {J.~L.}\ \bibnamefont
  {Jacobsen}}, \bibinfo {author} {\bibfnamefont {N.}~\bibnamefont {Read}}, \
  and\ \bibinfo {author} {\bibfnamefont {H.}~\bibnamefont {Saleur}},\
  }\bibfield  {title} {\enquote {\bibinfo {title} {Dense loops, supersymmetry,
  and goldstone phases in two dimensions},}\ }\href {\doibase
  10.1103/PhysRevLett.90.090601} {\bibfield  {journal} {\bibinfo  {journal}
  {Phys. Rev. Lett.}\ }\textbf {\bibinfo {volume} {90}},\ \bibinfo {pages}
  {090601} (\bibinfo {year} {2003})}\BibitemShut {NoStop}%
\bibitem [{\citenamefont {Read}\ and\ \citenamefont
  {Saleur}(2001)}]{READ2001409}%
  \BibitemOpen
  \bibfield  {author} {\bibinfo {author} {\bibfnamefont {N.}~\bibnamefont
  {Read}}\ and\ \bibinfo {author} {\bibfnamefont {H.}~\bibnamefont {Saleur}},\
  }\bibfield  {title} {\enquote {\bibinfo {title} {Exact spectra of conformal
  supersymmetric nonlinear sigma models in two dimensions},}\ }\href {\doibase
  https://doi.org/10.1016/S0550-3213(01)00395-9} {\bibfield  {journal}
  {\bibinfo  {journal} {Nuclear Physics B}\ }\textbf {\bibinfo {volume}
  {613}},\ \bibinfo {pages} {409 -- 444} (\bibinfo {year} {2001})}\BibitemShut
  {NoStop}%
\bibitem [{\citenamefont {Owczarek}\ and\ \citenamefont
  {Prellberg}(1995)}]{crossingloop}%
  \BibitemOpen
  \bibfield  {author} {\bibinfo {author} {\bibfnamefont {A.~L.}\ \bibnamefont
  {Owczarek}}\ and\ \bibinfo {author} {\bibfnamefont {T.}~\bibnamefont
  {Prellberg}},\ }\bibfield  {title} {\enquote {\bibinfo {title} {The collapse
  point of interacting trails in two dimensions from kinetic growth
  simulations},}\ }\href {\doibase 10.1007/BF02181210} {\bibfield  {journal}
  {\bibinfo  {journal} {Journal of Statistical Physics}\ }\textbf {\bibinfo
  {volume} {79}},\ \bibinfo {pages} {951--967} (\bibinfo {year}
  {1995})}\BibitemShut {NoStop}%
\bibitem [{\citenamefont {Ziff}\ \emph {et~al.}(1991)\citenamefont {Ziff},
  \citenamefont {Kong},\ and\ \citenamefont {Cohen}}]{PhysRevA.44.2410}%
  \BibitemOpen
  \bibfield  {author} {\bibinfo {author} {\bibfnamefont {Robert~M.}\
  \bibnamefont {Ziff}}, \bibinfo {author} {\bibfnamefont {X.~P.}\ \bibnamefont
  {Kong}}, \ and\ \bibinfo {author} {\bibfnamefont {E.~G.~D.}\ \bibnamefont
  {Cohen}},\ }\bibfield  {title} {\enquote {\bibinfo {title} {Lorentz
  lattice-gas and kinetic-walk model},}\ }\href {\doibase
  10.1103/PhysRevA.44.2410} {\bibfield  {journal} {\bibinfo  {journal} {Phys.
  Rev. A}\ }\textbf {\bibinfo {volume} {44}},\ \bibinfo {pages} {2410--2428}
  (\bibinfo {year} {1991})}\BibitemShut {NoStop}%
\bibitem [{\citenamefont {Martins}\ \emph {et~al.}(1998)\citenamefont
  {Martins}, \citenamefont {Nienhuis},\ and\ \citenamefont
  {Rietman}}]{PhysRevLett.81.504}%
  \BibitemOpen
  \bibfield  {author} {\bibinfo {author} {\bibfnamefont {M.~J.}\ \bibnamefont
  {Martins}}, \bibinfo {author} {\bibfnamefont {B.}~\bibnamefont {Nienhuis}}, \
  and\ \bibinfo {author} {\bibfnamefont {R.}~\bibnamefont {Rietman}},\
  }\bibfield  {title} {\enquote {\bibinfo {title} {Intersecting loop model as a
  solvable super spin chain},}\ }\href {\doibase 10.1103/PhysRevLett.81.504}
  {\bibfield  {journal} {\bibinfo  {journal} {Phys. Rev. Lett.}\ }\textbf
  {\bibinfo {volume} {81}},\ \bibinfo {pages} {504--507} (\bibinfo {year}
  {1998})}\BibitemShut {NoStop}%
\bibitem [{\citenamefont {Kager}\ and\ \citenamefont
  {Nienhuis}(2006)}]{Kager_2006}%
  \BibitemOpen
  \bibfield  {author} {\bibinfo {author} {\bibfnamefont {Wouter}\ \bibnamefont
  {Kager}}\ and\ \bibinfo {author} {\bibfnamefont {Bernard}\ \bibnamefont
  {Nienhuis}},\ }\bibfield  {title} {\enquote {\bibinfo {title} {Monte carlo
  study of the hull distribution for theq= 1 brauer model},}\ }\href {\doibase
  10.1088/1742-5468/2006/08/p08004} {\bibfield  {journal} {\bibinfo  {journal}
  {Journal of Statistical Mechanics: Theory and Experiment}\ }\textbf {\bibinfo
  {volume} {2006}},\ \bibinfo {pages} {P08004--P08004} (\bibinfo {year}
  {2006})}\BibitemShut {NoStop}%
\bibitem [{\citenamefont {Ikhlef}\ \emph {et~al.}(2007)\citenamefont {Ikhlef},
  \citenamefont {Jacobsen},\ and\ \citenamefont {Saleur}}]{Ikhlef_2007}%
  \BibitemOpen
  \bibfield  {author} {\bibinfo {author} {\bibfnamefont {Yacine}\ \bibnamefont
  {Ikhlef}}, \bibinfo {author} {\bibfnamefont {Jesper}\ \bibnamefont
  {Jacobsen}}, \ and\ \bibinfo {author} {\bibfnamefont {Hubert}\ \bibnamefont
  {Saleur}},\ }\bibfield  {title} {\enquote {\bibinfo {title} {Non-intersection
  exponents of fully packed trails on the square lattice},}\ }\href {\doibase
  10.1088/1742-5468/2007/05/p05005} {\bibfield  {journal} {\bibinfo  {journal}
  {Journal of Statistical Mechanics: Theory and Experiment}\ }\textbf {\bibinfo
  {volume} {2007}},\ \bibinfo {pages} {P05005--P05005} (\bibinfo {year}
  {2007})}\BibitemShut {NoStop}%
\bibitem [{\citenamefont {Vasseur}\ \emph {et~al.}(2019)\citenamefont
  {Vasseur}, \citenamefont {Potter}, \citenamefont {You},\ and\ \citenamefont
  {Ludwig}}]{vasseur}%
  \BibitemOpen
  \bibfield  {author} {\bibinfo {author} {\bibfnamefont {Romain}\ \bibnamefont
  {Vasseur}}, \bibinfo {author} {\bibfnamefont {Andrew~C.}\ \bibnamefont
  {Potter}}, \bibinfo {author} {\bibfnamefont {Yi-Zhuang}\ \bibnamefont {You}},
  \ and\ \bibinfo {author} {\bibfnamefont {Andreas W.~W.}\ \bibnamefont
  {Ludwig}},\ }\bibfield  {title} {\enquote {\bibinfo {title} {Entanglement
  transitions from holographic random tensor networks},}\ }\href {\doibase
  10.1103/PhysRevB.100.134203} {\bibfield  {journal} {\bibinfo  {journal}
  {Phys. Rev. B}\ }\textbf {\bibinfo {volume} {100}},\ \bibinfo {pages}
  {134203} (\bibinfo {year} {2019})}\BibitemShut {NoStop}%
\bibitem [{\citenamefont {Kitaev}(2006)}]{KITAEV20062}%
  \BibitemOpen
  \bibfield  {author} {\bibinfo {author} {\bibfnamefont {Alexei}\ \bibnamefont
  {Kitaev}},\ }\bibfield  {title} {\enquote {\bibinfo {title} {Anyons in an
  exactly solved model and beyond},}\ }\href {\doibase
  https://doi.org/10.1016/j.aop.2005.10.005} {\bibfield  {journal} {\bibinfo
  {journal} {Annals of Physics}\ }\textbf {\bibinfo {volume} {321}},\ \bibinfo
  {pages} {2 -- 111} (\bibinfo {year} {2006})},\ \bibinfo {note} {january
  Special Issue}\BibitemShut {NoStop}%
\bibitem [{\citenamefont {Lavasani}\ \emph {et~al.}(2020)\citenamefont
  {Lavasani}, \citenamefont {Alavirad},\ and\ \citenamefont
  {Barkeshli}}]{lavasani2020measurementinduced}%
  \BibitemOpen
  \bibfield  {author} {\bibinfo {author} {\bibfnamefont {Ali}\ \bibnamefont
  {Lavasani}}, \bibinfo {author} {\bibfnamefont {Yahya}\ \bibnamefont
  {Alavirad}}, \ and\ \bibinfo {author} {\bibfnamefont {Maissam}\ \bibnamefont
  {Barkeshli}},\ }\href@noop {} {\enquote {\bibinfo {title}
  {Measurement-induced topological entanglement transitions in symmetric random
  quantum circuits},}\ } (\bibinfo {year} {2020}),\ \Eprint
  {http://arxiv.org/abs/2004.07243} {arXiv:2004.07243 [quant-ph]} \BibitemShut
  {NoStop}%
\end{thebibliography}%

\section{Appendix A: Clifford Circuits with Symmetry}
    A stabilizer state over $N$ qubits $\ket{\psi_\mathcal{S}}$ is defined to be the unique simultaneous +1 eigen-state of a set of stabilizer $\mathcal{S}$:
    \beqa
        s \ket{\psi_\mathcal{S}} = \ket{\psi_\mathcal{S}}\ \ \forall s\in \mathcal{S}
    \eeqa
    Where $\mathcal{S}=\{s_1, ..., s_N\}$ is a set of mutually commuting and independent (under multiplication) Pauli string operators. The algorithm for obtaining entanglement entropy from $\mathcal{S}$ was introduced in \cite{PhysRevX.9.031009}
    
    Since any non-identity Pauli string operator $s\in\mathcal{S}$ has spectrum $\{1, -1\}$, $\frac{1}{2}(s+1)$ is a projector to the $s$ 's positive eigen-space. Further the density matrix of $\ket{\psi_\mathcal{S}}$ can be explicitly written as:
    \beqa
    \rho_{\mathcal{S}}=
        \ket{\psi_\mathcal{S}}\bra{\psi_\mathcal{S}}
        =\prod_{i}\left(\frac{1+s_i}{2}\right) = \frac{1}{2^N}\sum_{g\in \mathcal{G}}g
    \eeqa
    Here $\mathcal{G}_\mathcal{S}=\{s_1^{b_1},...,s_N^{b_N}| b_i\in\{0,1\}\ \forall i\}$ is the finite abelian group spanned by $\mathcal{S}$ under multiplication, named the stabilizer group of $\ket{\psi_\mathcal{S}}$. 
    
    A stabilizer state $\ket{\psi_\mathcal{S}}$ can be efficiently stored in memory by only keeping track of $\mathcal{S}$, which takes $\mathcal{O}(N^2)$ bytes. One can also obtain quantities involving $\ket{\psi_\mathcal{S}}$ by only referring to $\mathcal{S}$. A method for calculating bipartite entanglement entropy from $\mathcal{S}$ was introduced in \cite{PhysRevB.100.134306}. For the spin glass order parameter \ref{order}, the two-point correlation square term can be expressed as:
    \beqa
        &\langle \psi_{\mathcal{S}}|Z_i Z_j|\psi_{\mathcal{S}} \rangle ^2\\
        =& \Tr(\rho_{\mathcal{S}} Z_i Z_j \rho_{\mathcal{S}} Z_i Z_j)\\
        =& \frac{1}{2^{2N}}\Tr\left(\prod_k (1+s_k) Z_i Z_j \prod_l (1+s_l) Z_i Z_j\right)\\
        =& \frac{1}{2^{2N}}\Tr\left(\prod_k (1+s_k)(1+c_{ki}c_{kj} s_k)\right)\\
        =& \frac{1}{2^{2N}}\Tr\left(\prod_k (1+s_k)(1+c_{ki}c_{kj})\right)\\
        =& \frac{1}{2^{2N}}\Tr\left(\prod_k (1+s_k)\right)\prod_l (1+c_{li}c_{lj})\\
        =& \frac{1}{2^{N}}\prod_l (1+c_{li}c_{lj})\\
        =& \prod_l \mathbbm{1}[c_{li}c_{lj}=1]
    \eeqa
    where $\mathbf{c}$ is a $\{1, -1\}$ valued matrix such that $Z_i s_k = c_{ki} s_k Z_i$. The one-point square term can similarly be obtained as:
    \beqa
        \langle \psi_{\mathcal{S}}|Z_i |\psi_{\mathcal{S}} \rangle ^2 = \prod_l \mathbbm{1}[c_{li}=1]
    \eeqa
    Clifford gates over $N$ qubits $\mathcal{C}_n$ is a class of unitary gates with the property of always mapping one Pauli string operator to another. The action of Clifford gate $U\in \mathcal{C}_N$ on a stabilizer state $\ket{\psi_\mathcal{S}}$ is given by:
    
    \beqa
        U^\dagger \ket{\psi_\mathcal{S}}\bra{\psi_\mathcal{S}}U = \prod_{i}\left(\frac{1+U^\dagger s_i U}{2}\right) = \ket{\psi_{\mathcal{S}^U}}\bra{\psi_{\mathcal{S}^U}}
    \eeqa
    where $\mathcal{S}^U=\{s_1^U,..., s_N^ U\}=\{U^\dagger s_1 U,..., U^\dagger s_N U\}$ is still a valid set of stabilizers. So Clifford group also leaves the set of stabilizer states invariant. 
    
    A $N$-qubit Clifford gate is completely decided by its action on single site Pauli operators $\{X_i, Z_i\}_{i\in [N]}$. Clearly the mapping must preserve the commutation relation within $\{X_i, Z_i\}_{i\in [N]}$. Moreover, it can be shown that any mapping that maps $\{X_i, Z_i\}_{i\in [N]}$ to the set of Pauli string operators and preserves their commutation relations uniquely (up to a phase factor) determines a $U\in C_N$.
    
    In the maintext we focused on a subset of $\mathcal{C}_N$ that respects the Ising symmetry, namely the $\mathbb{Z}_2$ symmetric Clifford gates $\mathcal{C}_N^{\text{sym}}$. Such gates can be characterized by their defining property of leaving the global flipping operator $T=\prod_i X_i$ invariant:
    \beqa
        \mathcal{C}_N^{\text{sym}}=\{U\in \mathcal{C}_N|U^\dagger T U = T\}
    \eeqa
    
    Similar to generic Clifford gates, $\mathcal{C}_N^{\text{sym}}$ as a finite discrete group can be generated by a much smaller set of one- and two-qubit gates. The description of this set is more clear in the Majorana picture through the Jordan-Wigner transformation:
    \beqa
        \gamma_{2i-1}&=(\prod_{j<i}X_j)Y_i\\
        \gamma_{2i}&=(\prod_{j<i}X_j)Z_i
    \eeqa
    
    Because the transformation always maps a Pauli string operator to a Majorana one and vise versa, we can conclude that in the Majorana picture a Clifford gate always maps one Majorana string operator to another (up to some phase factor). The $\mathbb{Z}_2$ symmetry constraint guarantees that the action of $U\in\mathcal{C}_N^{\text{sym}}$ preserves the Majorana parity, and is local in both spin and Majorana picture.
    
    Within the Majorana picture, $\mathcal{C}_N^{\text{sym}}$ is generated by two kinds of gates: the two-Majorana swap gate $U^s=\exp(\frac{\pi}{4 }\gamma_{1}\gamma_{2})$:
    \beqa
        (U^s)^\dagger\ \gamma_1\  U^s &= \gamma_{2}\\
        (U^s)^\dagger\ \gamma_{2}\  U^s &= -\gamma_1
    \eeqa
    and the four-Majorana ``parity gate'' (acting like a multiplication by the local fermion parity operator) $U^p = \exp(\frac{i\pi}{4}\gamma_1\gamma_2\gamma_3\gamma_4)$:
    \beqa
        (U^p)^\dagger\ \gamma_1\  U^p &= i\gamma_{2}\gamma_{3}\gamma_{4}\\
        (U^p)^\dagger\ \gamma_{2}\  U^p &= -i\gamma_{1}\gamma_{3}\gamma_{4}\\
        (U^p)^\dagger\ \gamma_{3}\  U^p &= i\gamma_{1}\gamma_{2}\gamma_{4}\\
        (U^p)^\dagger\ \gamma_{4}\  U^p &= -i\gamma_{1}\gamma_{2}\gamma_{3}
    \eeqa
    
    To numerically sample an element $U$ from $\mathcal{C}_2^{\text{sym}}$, first a random element is picked in $\mathcal{P}_2 - \{I_1 I_2, X_1 X_2\}$ as $X_1^U$ ($\mathcal{P}_2$ is the set of 2-Pauli operators), then $X_2^U$ is automatically determined through $X_1^U X_2^U = X_1 X_2$. $Z_1$ is sampled from a subset of $\mathcal{P}_2 - \{I_1 I_2, X_1 X_2, X_1^U, X_2^U\}$ that commutes with $X_2^U$ and anti-commutes with $X_1^U$. Finally $Z_2^U$ can be sampled in a similar manner.
    
\section{Appendix B: Details of sampling procedure}
    In the main text, we are mainly concerned about properties of the ensemble of late time steady states produced by some given circuit architecture. In this section we explain how we sample states from this ensemble numerically. 
    
\begin{figure}
        \includegraphics[width=\linewidth]{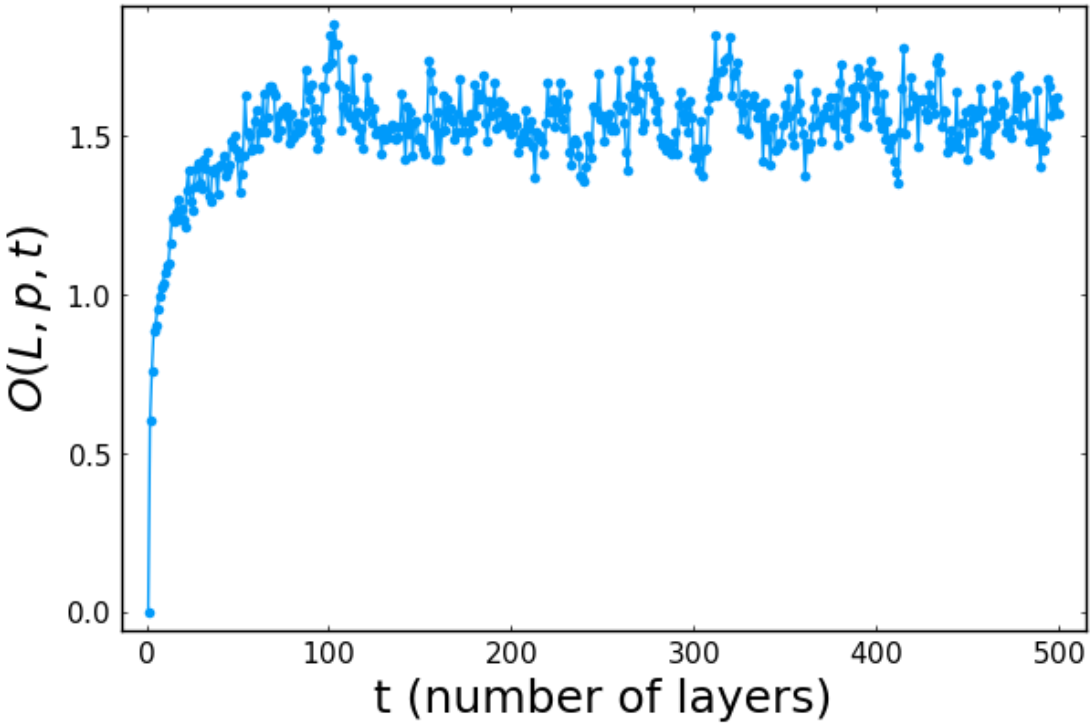}
        \caption{Spin glass order parameter $O$ as a function of time for a system of 512 spins at $r=1, p=p_c=0.39$}
        \label{transient}
    \end{figure}

    For a given random realization of circuit with size $L$, we first evolve the initial state (which is typically chosen to be a product state) for $\tau L$ steps so that it reaches the equilibrium, then sample the evolving state every $\Delta t$ steps. By increasing $\Delta t$, one can reduce the correlation between two adjacent sampled states and increase the convergence speed of target quantities. In our simulations $\Delta t$ is fixed to be $32$. The selection of $\tau$ is usually simulation-wise as $\tau L$ needs to be larger than the time required for the system to reach equilibrium, and the latter is usually architecture and parameter dependent. To decide $\tau$ one can plot the quantity of interest as a function of time steps then take any time after which the quantity saturates divided by $L$ as $\tau$. As an example, FIG.\ref{transient} shows the transient behavior of $O(L=512, p=0.39, t)$ at $r=1$. 
    
\section{Appendix C: Entanglement entropy on other cross sections}

\begin{figure}
    \includegraphics[width=\linewidth]{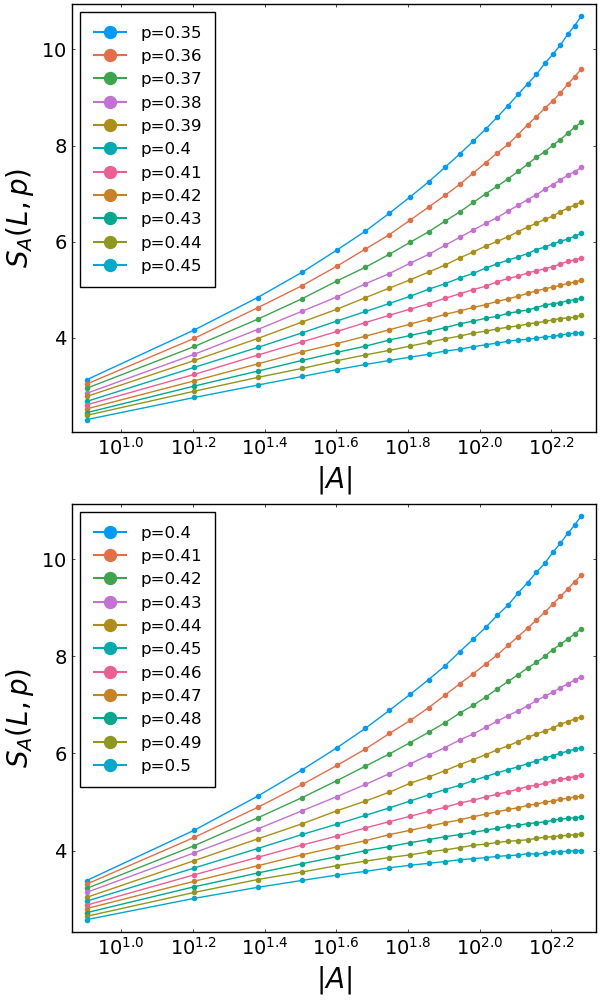}
    \caption{Entanglement entropy versus log of partition size near the critical points at $r=0.25$ (top) and $r=0.75$ (bottom). $L=768$.}
    \label{appendixB}
\end{figure}

\begin{figure}[H]
    \includegraphics[width=\linewidth]{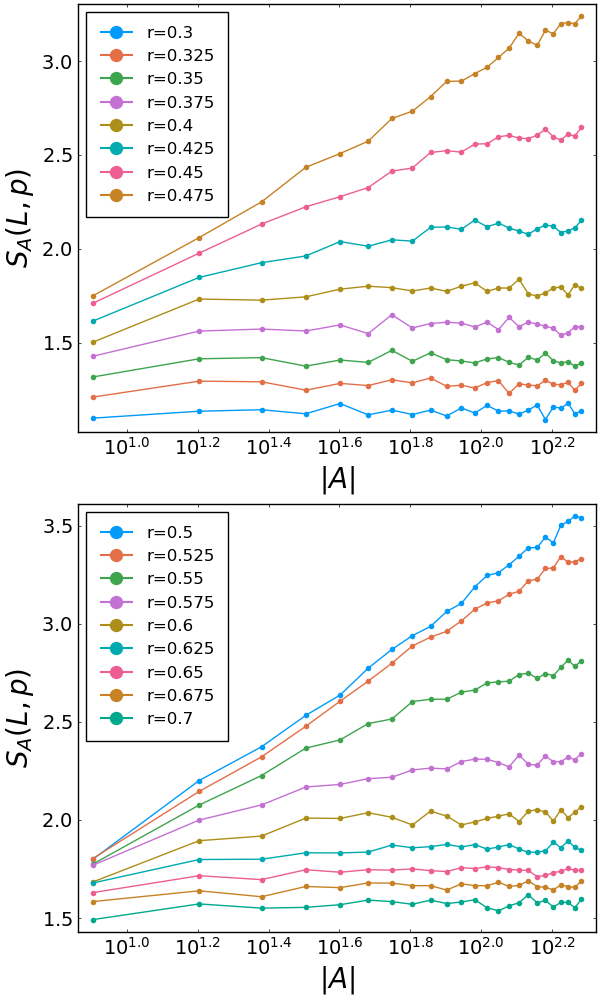}
    \caption{Entanglement entropy versus log of partition size near the critical points at $p = 0.75$, two figures for two different ranges of $r$. $L=768$.}
    \label{appendixB_2}
\end{figure}

We obtain entanglement entropy scaling for three additional cross sections of the phase diagram: $r=0.25,0.75$ (Fig. \ref{appendixB}) and $p=0.75$ (Fig. \ref{appendixB_2}). 

\section{Appendix D: Data for Alternative Architecture}

\begin{figure}[H]
    \includegraphics[width=\linewidth]{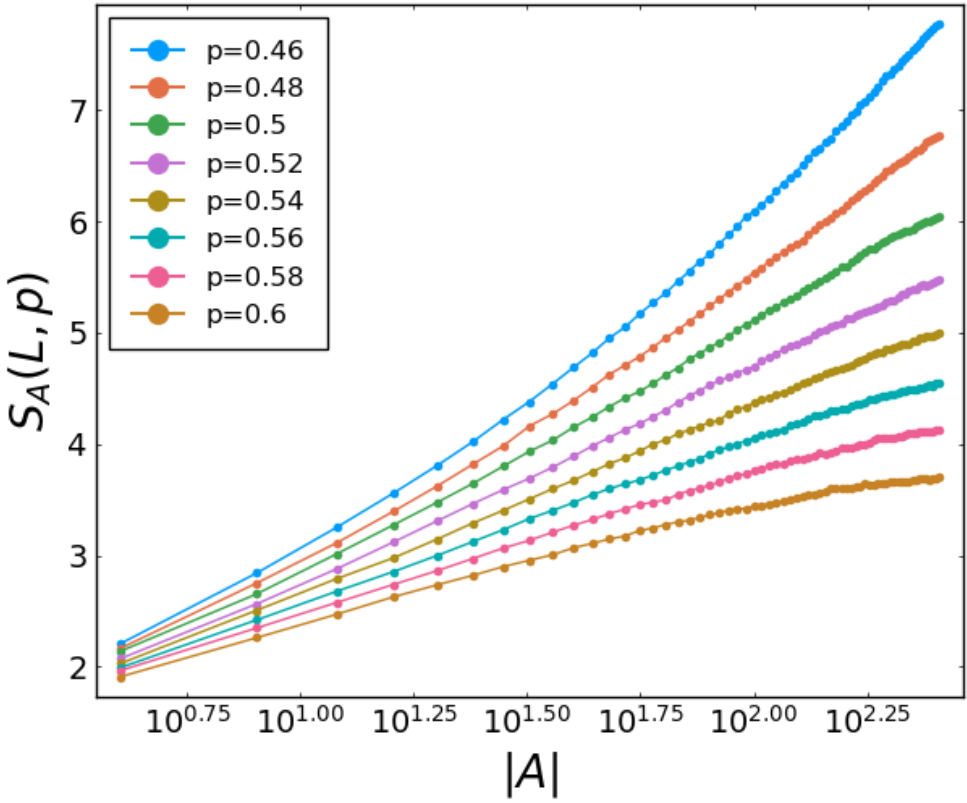}
    \caption{Entanglement entropy versus log of partition size, for various $p$, for the alternative architecture.}
    \label{arch2data1}
\end{figure}

\begin{figure}[H]
    \includegraphics[width=\linewidth]{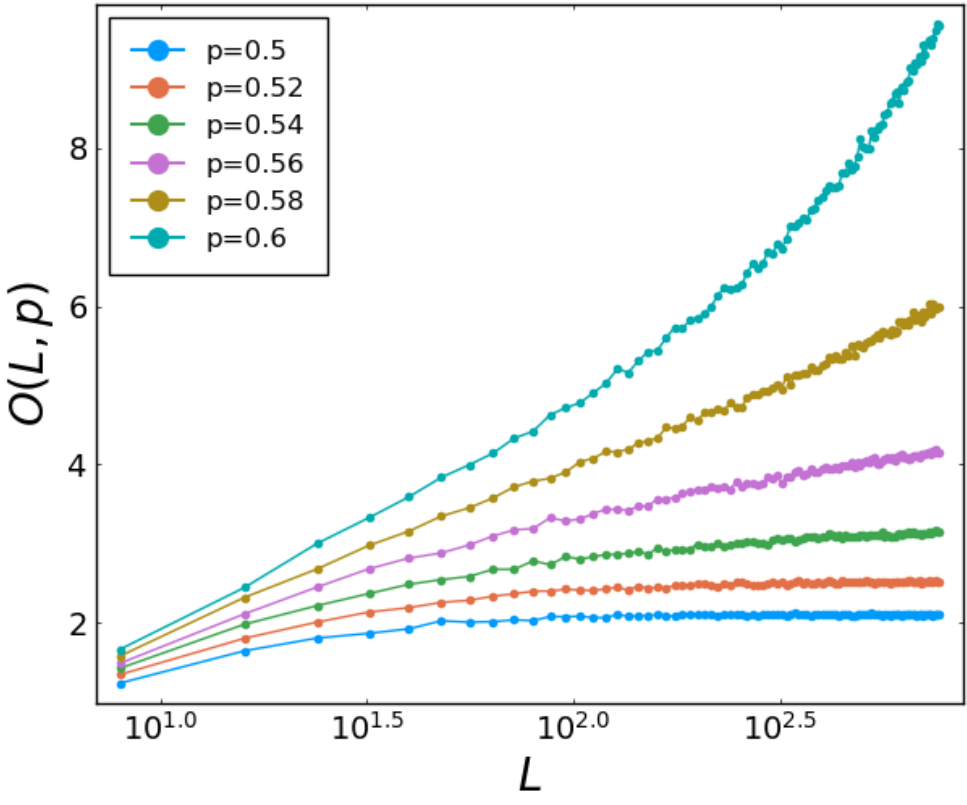}
    \caption{Spin glass order parameter versus log of system size, for various $p$, for the alternative architecture.}
    \label{arch2data2}
\end{figure}

For the alternative architecture (Fig. \ref{arch2} in main text), we present the entanglement entropy scaling and spin glass order parameter data in Figures \ref{arch2data1} and \ref{arch2data2}.  Based on the data, we conclude that the volume law phase is destroyed at $p_{c,S}\approx 0.52$ and the spin glass order onsets at $p_{c,O}\approx 0.58$.  The intermediate regime has spin-glass correlation and entanglement scaling that is neither clearly area or volume law.  These could be due to finite size effects and require larger systems for further study.

\section{Appendix E: Further data for (2+1)D circuit at $p=0.3$}

\begin{figure}[H]
    \includegraphics[width=\linewidth]{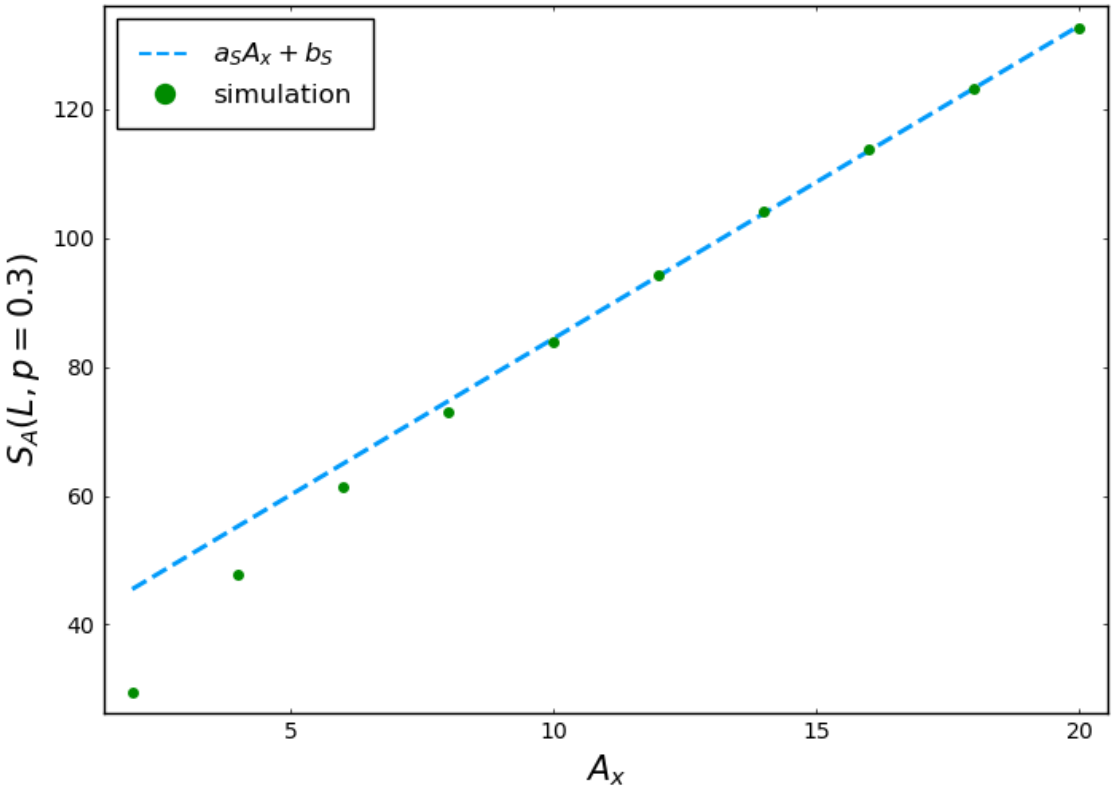}
    \caption{Simulated bipartite entanglement entropy $S_{A}(L, p)$ at $p=0.3$ in (2+1)D circuit with varying partition size $A=(A_x, A_y)$. The total system size $L=(L_x, L_y)=(60, 20)$ and $A_y = 20$ are fixed while $A_x$ is varying. For dashed line $(a_S, b_S) = (4.87, 35.85)$}
    \label{s_at_03}
\end{figure}

\begin{figure}[H]
    \includegraphics[width=\linewidth]{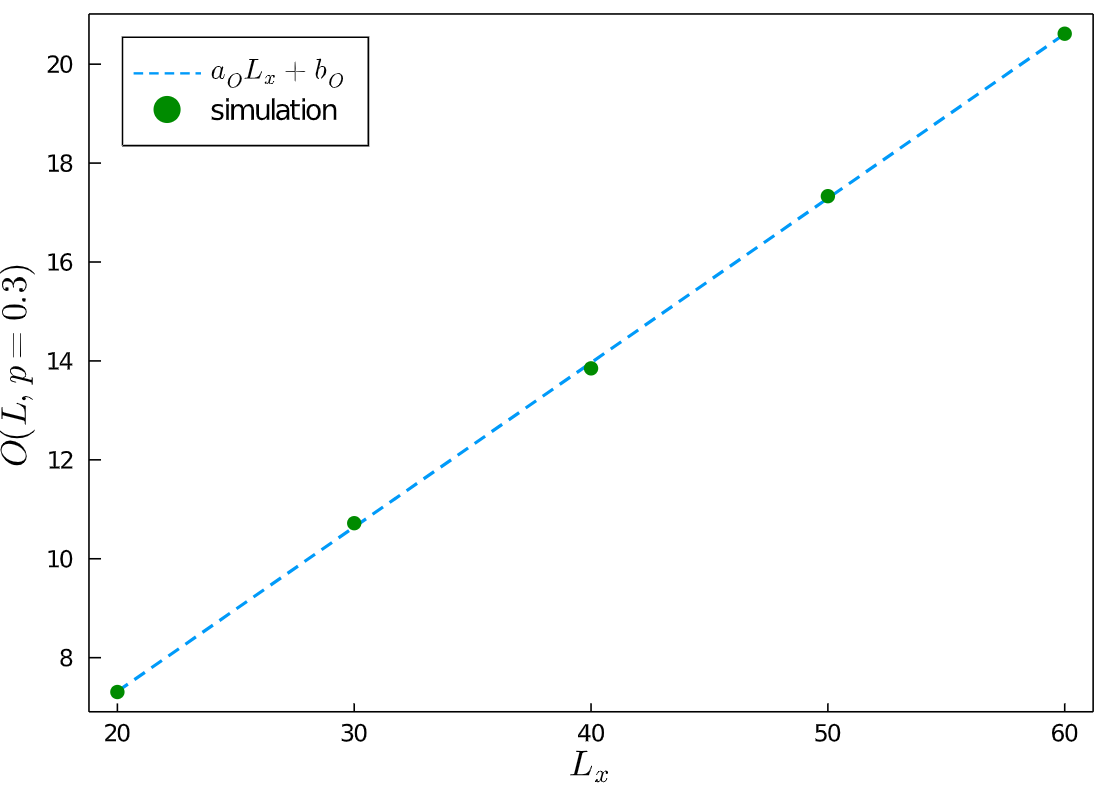}
    \caption{Simulated spin-glass order parameter $O(L, p)$ at $p=0.3$ in (2+1)D circuit with $L_y = 20$ fixed and $L_x$ varying. For dashed line $(a_O, b_O) = (0.33, 0.67)$}
    \label{o_at_03}
\end{figure}

Fig.\ref{s_at_03} and fig.\ref{o_at_03} present the behaviors of $S_{A}(L,p)$ and $O(L,p)$ at $p=0.3$ in (2+1)D circuit (see fig.\ref{2d} in maintext). Our result shows that $S_{A}(L,p)$ / $O(L,p)$ scales linearly with partition size / system size when $p=0.3$.

\end{document}